\documentclass[useAMS,usenatbib]{mn2e}
\usepackage{graphicx}
\usepackage{amssymb}
\usepackage{lscape}
\usepackage{ulem}
\usepackage{txfonts}

\def\dv{$\Delta V$}
\def\kms{km~s$^{-1}$}
\def\cm{cm$^{-2}$}
\def\lya{Ly$\alpha$}
\def\nhi{$N$(H\,{\sc i})}
\def\hi{H\,{\sc i}}
\def\si2{Si\,{\sc ii}}
\def\c4{C\,{\sc iv}}
\def\mg2{Mg\,{\sc ii}}
\def\n5{N\,{\sc v}}
\def\fe2{Fe\,{\sc ii}}
\def\al2{Al\,{\sc ii}}
\def\zn2{Zn\,{\sc ii}}
\def\c2s{C\,{\sc ii}$^{\star}$}

\def\ebmv{E$(B-V)$}

\title[Proximate DLAs in the SDSS]{Metallicities and dust content of
proximate damped Lyman alpha systems in the Sloan Digital Sky Survey}

\author[Ellison et al.] {Sara L. Ellison$^1$,  J. Xavier Prochaska$^2$,
J. Trevor Mendel$^1$\\
$^1$ Department of Physics and Astronomy, University of Victoria, 
Victoria, British Columbia, V8P 1A1, Canada.\\
$^2$ Department of Astronomy and Astrophysics, UCO/Lick Observatory, University of California, 1156 High Street, Santa Cruz, CA 95064, USA 
}

\begin{document}

\maketitle

\begin{abstract}

Composite spectra of 85 proximate absorbers (log \nhi\ $\ge$ 20 \cm\
and velocity difference between the absorption and emission redshift,
\dv\ $<$ 10,000 \kms) in the Sloan Digital Sky Survey are used to
investigate the trends of metal line strengths with velocity
separation from the QSO.  We construct composites in 3 velocity bins:
\dv\ $<$ 3000 \kms, 3000 $<$ \dv\ $<$ 6000 \kms\ and \dv\ $>$ 6000
\kms, with further sub-samples to investigate the metal line
dependence on \nhi\ and QSO luminosity.  Low (e.g. SiII and FeII) and
high ionization (e.g. SiIV and CIV) species alike have equivalent
widths (EWs) that are larger by factors of 1.5 -- 3 in the \dv\ $<$
3000 \kms\ composite, compared to the \dv\ $>$ 6000 \kms\ spectrum.
The EWs show an even stronger dependence on \dv\ if only the highest
neutral hydrogen column density (log \nhi\ $\ge$ 20.7) absorbers are
considered.  We conclude that PDLAs generally have higher
metallicities than intervening absorbers, with the enhancement being a
function of both \dv\ and \nhi.  It is also found that absorbers near
QSOs with lower rest-frame UV luminosities have significantly stronger
metal lines.  We speculate that absorbers near to high luminosity QSOs
may have had their star formation prematurely quenched.  Finally, we
search for the signature of dust reddening by the PDLAs, based on an
analysis of the QSO continuum slopes relative to a control sample and
determine a limit of \ebmv\ $<$ 0.014 for an SMC extinction curve.
This work provides an empirical motivation for distinguishing between
proximate and intervening DLAs, and establishes a connection between
the QSO environment and galaxy properties at high redshifts.

\end{abstract}


\section{Introduction}

Damped Lyman alpha (DLA) systems with small velocity separations
(typically \dv\ $<$3000 to 5000 \kms) from the systemic redshift of the
background QSO are traditionally excluded by most statistical surveys
(e.g.  Lanzetta et al. 1991, 1995; Wolfe et al. 1995; Storrie-Lombardi
et al. 1996; Ellison et al. 2001; Peroux et al. 2001; Jorgenson et
al. 2006; Ellison et al. 2008; Prochaska \& Wolfe 2009; Noterdaeme et
al. 2009).  The logic for the \dv\ criterion was originally imposed to
avoid absorbers that were not representative of the intervening
population (e.g. those clustered around the QSO), or directly
associated with the QSO host or its outflows.  A further concern for
small \dv\ absorbers is that proximity to the intense QSO radiation
field may alter the internal ionization balance of the galactic
interstellar medium (ISM).  In turn, this could lead to the
requirement for complex ionization corrections when converting
observed column densities into elemental abundances.  Despite some
notable exceptions (Prochaska et al. 2002a,b; Dessauges-Zavadsky et
al. 2004, 2006; Milutinovic et al. 2010), ionization corrections in
intervening DLAS can generally be ignored, even when the column
density is relatively low.

Although these proximity concerns have driven DLA selection criteria
for two decades, there is relatively little observational evidence to
support the imposition of a velocity cut.  Apart from a higher
incidence by a factor of $\sim$ 2--4 (Ellison et al. 2002; Russell,
Ellison \& Benn 2006; Prochaska, Hennawi \& Herbert-Fort 2008), the
only empirical evidence that the proximate DLAs (PDLAs) differ from
the intervening absorbers has been the higher incidence of \lya\
emission detected in the DLA trough (M\o ller \& Warren 1993; M\o
ller, Warren \& Fynbo 1998; Hennawi et al. 2009).  The QSO's radiation
does not manifest itself in gross differences in highly ionized ISM
species.  For example, Fox et al. (2009) found no evidence for a more
frequent occurence of NV.  The CIV and OVI column densities of the
proximate absorbers are also indistinguishable from the intervening
population (Fox et al. 2007a, 2007b), although larger samples are needed
to confirm this.  It has also been concluded by Rix
et al. (2007) (based on the few PDLAs with high resolution echelle
spectra) that their chemical abundance patterns and overall metal
enrichment are consistent with the intervening population.

We have recently re-assessed the chemical properties of PDLAs using a
sample of 16 absorbers with \dv\ $<$ 3000 \kms\ using high resolution
echelle data.  In Ellison et al. (2010) we reported the first evidence
that some PDLAs do exhibit both distinct metal enrichment and evidence
for ionization by a hard radiation source.  Specifically, it was found
that although the proximate absorbers can indeed display a wide range
of metallicities, at high \nhi\ they are typically a factor of 3 more
metal-rich than the intervening population.  Although many of the
proximate absorbers exhibit NV (including absorption with large
velocity offsets), the most intriguing evidence for a hard radiation
source came from non-solar abundances of S and Ar.  Expansion of the
PDLA data set is clearly required to confirm these initial findings
and to understand the case-by-case dependence on QSO proximity.

In this paper, we take a complementary approach by leveraging the
large statistical power of the Sloan Digital Sky Survey (SDSS).
Although the majority of DLA abundance studies are performed at
resolutions $R \sim$ 40,000, with high enough S/N it is possible to
detect even weak, unsaturated lines in spectra whose resolution is an
order of magnitude lower (e.g. Pettini et al. 1994).  Our strategy
here is to construct high S/N composite spectra with various criteria
(e.g. \dv, \nhi, QSO luminosity), in order to investigate which
quantities most significantly affect the metallicities of the PDLAs.

Unless otherwise stated, we assume $\Omega_M = 0.3$, $\Omega_{\Lambda}=0.7$
and H$_0$ = 70 \kms\ Mpc$^{-1}$.

\section{Sample selection}\label{sample_sec}

\begin{center}
\begin{table*}
\begin{tabular}{lcccccccr}
\hline 
QSO                 &  $i$ mag &  Plate& MJD   & Fibre& $z_{\rm em}$ &  $z_{\rm abs}$ & log \nhi\ & \dv\ (\kms)\\ \hline
J012747.80+140543.2 & 18.56 & 0425 & 51898 & 278 & 2.4877 & 2.4416 & 20.30 & 3991.57 \\
J014214.74+002324.3 & 18.08 & 0401 & 51788 & 500 & 3.3734 & 3.3481 & 20.40 & 1733.61 \\
J023903.43-003850.8 & 18.71 & 0408 & 51821 & 134 & 3.0782 & 3.0185 & 20.35 & 4371.54 \\
J025518.58+004847.6 & 18.87 & 0410 & 51816 & 466 & 3.9936 & 3.9145 & 21.40 & 4771.41 \\
J073718.16+323631.5 & 19.10 & 0541 & 51959 & 151 & 3.0175 & 2.8926 & 20.30 & 9487.00 \\
J075901.28+284703.4 & 19.02 & 0859 & 52317 & 453 & 2.8550 & 2.8223 & 21.05 & 2532.00 \\
J080050.28+192058.9 & 18.89 & 1922 & 53315 & 152 & 3.9533 & 3.9465 & 20.25 & 448.52  \\
J081240.68+320808.6 & 17.43 & 0861 & 52318 & 333 & 2.7045 & 2.6259 & 21.30 & 6391.40 \\
J081256.05+563746.9 & 18.80 & 1872 & 53386 & 474 & 3.3170 & 3.2251 & 20.00 & 6425.95 \\
J081256.05+563746.9 & 18.80 & 1872 & 53386 & 474 & 3.3170 & 3.3387 & 20.35 & -1310.54\\
J081518.56+291153.9 & 19.33 & 0930 & 52618 & 593 & 4.2552 & 4.2588 & 20.50 & -205.44 \\
J082107.61+310751.2 & 17.04 & 0931 & 52619 & 491 & 2.6193 & 2.5347 & 20.15 & 7060.41 \\
J082531.88+263619.2 & 18.93 & 1267 & 52932 & 051 & 2.5640 & 2.5409 & 20.45 & 1933.80 \\
J082612.54+451355.7 & 19.16 & 0548 & 51986 & 167 & 3.8170 & 3.7068 & 20.50 & 6979.90 \\
J082638.59+515233.2 & 17.02 & 0442 & 51882 & 528 & 2.8438 & 2.8333 & 20.80 & 820.62  \\
J083510.92+065052.8 & 18.37 & 1297 & 52963 & 517 & 3.9781 & 3.9556 & 20.35 & 1359.00 \\
J083914.14+485125.7 & 18.64 & 0550 & 51959 & 460 & 2.9673 & 2.9692 & 20.60 & -136.08 \\
J090017.61+490001.9 & 18.72 & 0765 & 52254 & 306 & 3.2080 & 3.2054 & 20.95 & 185.42  \\
J090033.49+421546.8 & 16.78 & 0831 & 52294 & 201 & 3.2954 & 3.2456 & 20.30 & 3477.11 \\
J090940.67+330347.6 & 18.51 & 1272 & 52989 & 022 & 3.7835 & 3.6583 & 20.60 & 7922.54 \\
J091210.35+054742.0 & 18.09 & 1194 & 52703 & 257 & 3.2407 & 3.1231 & 20.35 & 8398.34 \\
J091223.02+562128.5 & 18.75 & 0451 & 51908 & 312 & 2.9816 & 2.8894 & 20.55 & 7027.37 \\
J091548.90+302542.6 & 18.85 & 1936 & 53330 & 614 & 3.1928 & 3.0678 & 20.45 & 9091.90 \\
J092914.49+282529.1 & 17.54 & 1940 & 53383 & 441 & 3.4044 & 3.2629 & 21.10 & 9743.62 \\
J093019.58+423803.9 & 18.86 & 0870 & 52325 & 042 & 3.7363 & 3.6098 & 20.20 & 8119.55 \\
J094453.89+372840.2 & 18.81 & 1276 & 53035 & 172 & 3.3397 & 3.3356 & 20.45 & 262.81  \\
J095256.41+332939.0 & 18.74 & 1945 & 53387 & 032 & 3.3962 & 3.3468 & 20.70 & 3369.33 \\
J095744.46+330820.7 & 18.15 & 1948 & 53388 & 137 & 4.2088 & 4.1794 & 20.45 & 2039.99 \\
J095817.81+494618.3 & 18.64 & 1006 & 52708 & 474 & 2.3555 & 2.2909 & 20.65 & 5831.18 \\
J095937.11+131215.4 & 17.43 & 1744 & 53055 & 368 & 4.0717 & 3.9131 & 20.15 & 9540.28 \\
J100409.35+120256.5 & 18.87 & 1744 & 53055 & 109 & 2.8804 & 2.8006 & 20.60 & 6232.89 \\
J101725.88+611627.5 & 18.13 & 0771 & 52370 & 155 & 2.8069 & 2.7681 & 20.60 & 3073.19 \\
J102611.35+341459.9 & 18.45 & 1958 & 53385 & 449 & 3.3991 & 3.4141 & 20.60 & -1021.19\\
J102619.09+613628.8 & 18.43 & 0772 & 52375 & 227 & 3.8442 & 3.7853 & 20.35 & 3657.30 \\
J103403.87+380248.4 & 18.15 & 1998 & 53433 & 343 & 3.5605 & 3.5161 & 20.70 & 2948.23 \\
J104837.40-002813.6 & 19.28 & 0276 & 51909 & 313 & 4.0070 & 3.8880 & 20.65 & 7214.72 \\
J105123.03+354534.3 & 18.54 & 2090 & 53463 & 105 & 4.8990 & 4.8208 & 20.40 & 4003.30 \\
J110855.47+120953.3 & 18.62 & 1604 & 53078 & 383 & 3.6737 & 3.5453 & 20.75 & 8302.28 \\
J111151.60+133235.9 & 17.15 & 1752 & 53379 & 242 & 2.4328 & 2.3820 & 20.30 & 4401.43 \\
J111611.73+411821.5 & 18.04 & 1439 & 53003 & 595 & 2.9829 & 2.9426 & 20.20 & 3012.79 \\
J113002.34+115438.3 & 18.39 & 1606 & 53055 & 102 & 3.3939 & 3.3171 & 20.20 & 5226.94 \\
J113008.19+535419.8 & 17.60 & 1014 & 52707 & 211 & 3.0495 & 2.9870 & 20.25 & 4575.67 \\
J113130.41+604420.7 & 17.60 & 0776 & 52319 & 250 & 2.9069 & 2.8755 & 20.50 & 2428.55 \\
J113354.89+022420.9 & 18.92 & 0513 & 51989 & 428 & 3.9865 & 3.9147 & 20.65 & 4332.45 \\
J115526.34+351052.6 & 18.85 & 2035 & 53436 & 011 & 2.8374 & 2.7582 & 20.70 & 6247.59 \\
J120359.07+341114.2 & 18.58 & 2099 & 53469 & 635 & 3.7484 & 3.6740 & 20.60 & 4718.10 \\
J120359.07+341114.2 & 18.58 & 2099 & 53469 & 635 & 3.7484 & 3.6860 & 20.60 & 3917.08 \\
J121324.58+423538.5 & 18.85 & 1450&  53120&  455&  3.7663&  3.7646&  20.60&  56.65  \\
J122040.23+092326.8 & 18.05 & 1230 & 52672 & 066 & 3.1462 & 3.1322 & 20.75 & 978.39  \\
J123840.93+343703.3 & 18.44 & 2020 & 53431 & 490 & 2.5718 & 2.4712 & 20.80 & 8551.19 \\
J123937.56+343701.8 & 18.41 & 2020 & 53431 & 501 & 2.4632 & 2.4812 & 21.10 & -1727.50\\
J124138.32+461717.0 & 18.57 & 1455 & 53089 & 257 & 2.7697 & 2.6672 & 20.70 & 8300.70 \\
J124640.37+111302.9 & 18.02 & 1694 & 53472 & 085 & 3.1475 & 3.0975 & 20.45 & 3594.52 \\
J125659.47+301439.0 & 18.71 & 2011 & 53499 & 388 & 2.9457 & 2.8805 & 20.70 & 4967.33 \\
J125759.22-011130.2 & 18.64 & 0293 & 51994 & 127 & 4.1120 & 4.0208 & 20.30 & 5358.04 \\
J125832.14+290903.1 & 18.95 & 2011 & 53499 & 171 & 3.4830 & 3.3512 & 20.55 & 8928.92 \\
J130152.57-030729.3 & 18.47 & 0339 & 51692 & 277 & 3.0706 & 3.0510 & 20.50 & 1336.92 \\
J130426.15+120245.5 & 17.93 & 1696 & 53116 & 278 & 2.9775 & 2.9135 & 20.55 & 4835.33 \\
J130426.15+120245.5 & 17.93 & 1696 & 53116 & 278 & 2.9775 & 2.9289 & 20.35 & 3726.19 \\
J132235.12+584124.6 & 18.63 & 0959 & 52411 & 237 & 2.8507 & 2.8183 & 20.50 & 2511.27 \\
\hline 
\end{tabular}
\caption{\label{PDLA_tab} 85 PDLAs used in this study.  Magnitudes are
in the Petrosian system and are corrected for Galactic extinction.}
\end{table*}
\end{center}

\addtocounter{table}{-1}
\begin{center}
\begin{table*}
\begin{tabular}{lcccccccc}
\hline 
QSO                  &  $i$ mag &  Plate& MJD   & Fibre& $z_{\rm em}$ &  $z_{\rm abs}$ & log \nhi\ & \dv\ (\kms)\\ \hline
J133724.69+315254.5 & 18.54 & 2109 & 53468 & 052 & 3.1829 & 3.1742 & 21.30 & 617.43  \\
J135305.18-025018.2 & 18.70 & 0914 & 52721 & 235 & 2.4149 & 2.3618 & 20.30 & 4647.59 \\
J135317.11+532825.5 & 18.29 & 1043 & 52465 & 492 & 2.9153 & 2.8346 & 20.80 & 6192.41 \\
J135803.97+034936.0 & 18.62 & 0856 & 52339 & 196 & 2.8937 & 2.8535 & 20.40 & 3113.30 \\
J135842.92+652236.6 & 18.70 & 0498 & 51984 & 398 & 3.1731 & 3.0641 & 20.30 & 7576.96 \\
J135936.21+483923.1 & 18.44 & 1671 & 53446 & 305 & 2.3641 & 2.2572 & 20.40 & 9647.59 \\
J141906.32+592312.3 & 17.69 & 0789 & 52342 & 368 & 2.3189 & 2.2467 & 20.95 & 6597.23 \\
J141950.54+082948.2 & 18.66 & 1811 & 53533 & 198 & 3.0344 & 3.0504 & 20.40 & -1157.78\\
J142446.29+054717.3 & 17.91 & 1827 & 53531 & 269 & 2.3944 & 2.3072 & 20.40 & 7751.38 \\
J144127.66+475048.8 & 18.80 & 1674 & 53464 & 608 & 3.1977 & 3.2130 & 21.05 & -1091.46\\
J145329.53+002357.5 & 18.49 & 0309 & 51994 & 423 & 2.5417 & 2.4436 & 20.40 & 8389.79 \\
J145429.65+004121.2 & 19.42 & 0309 & 51994 & 466 & 2.6567 & 2.5657 & 20.15 & 7667.94 \\
J150726.32+440649.2 & 17.80 & 1677 & 53148 & 424 & 3.1133 & 3.0642 & 20.75 & 3595.06 \\
J152413.35+430537.4 & 18.86 & 1678 & 53433 & 614 & 3.9196 & 3.8809 & 20.70 & 2399.96 \\
J154153.46+315329.4 & 17.53 & 1581 & 53149 & 220 & 2.5530 & 2.4434 & 20.85 & 9388.11 \\
J155814.51+405337.0 & 18.85 & 1054 & 52516 & 602 & 2.6403 & 2.5521 & 20.30 & 7356.66 \\
J160413.97+395121.9 & 17.95 & 1055 & 52761 & 459 & 3.1542 & 3.1625 & 21.75 & -656.45 \\
J164716.62+313254.4 & 18.53 & 1341 & 52786 & 025 & 2.5173 & 2.4936 & 20.25 & 2028.25 \\
J170353.98+362439.6 & 18.91 & 0820 & 52438 & 631 & 2.4764 & 2.4186 & 20.25 & 5046.92 \\
J173935.27+575201.7 & 19.06 & 0358 & 51818 & 605 & 3.2081 & 3.1357 & 20.35 & 5242.13 \\
J210025.03-064146.0 & 18.21 & 0637 & 52174 & 370 & 3.1295 & 3.0918 & 21.05 & 2729.34 \\
J211443.94-005532.7 & 18.68 & 0986 & 52443 & 185 & 3.4245 & 3.4420 & 20.50 & -1211.24\\
J223843.56+001647.9 & 19.02 & 0377 & 52145 & 560 & 3.4499 & 3.3654 & 20.40 & 5750.84 \\
J231543.56+145606.4 & 18.58 & 0745 & 52258 & 355 & 3.3769 & 3.2735 & 20.30 & 7149.85 \\
J232115.48+142131.5 & 18.44 & 0745 & 52258 & 232 & 2.5539 & 2.5729 & 20.60 & -1607.98\\
\hline 
\end{tabular}
\caption{Continued. 85 PDLAs used in this study.}
\end{table*}
\end{center}

\begin{figure}
\centerline{\rotatebox{270}{\resizebox{6.5cm}{!}
{\includegraphics{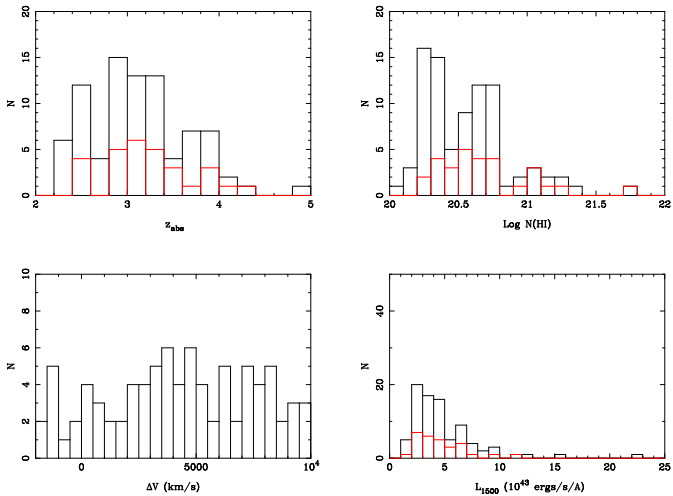}}}}
\caption{\label{data_histo}Distribution of absorber (absorption
redshift, \hi\ column density and velocity offset from the QSO
systemic redshift) and QSO (luminosity) properties measured from the
SDSS spectra for our sample of 85 PDLAs. }
\end{figure}

\subsection{PDLA sample}

The largest compilation of PDLAs is the SDSS data release (DR) 5
sample of Prochaska et al. (2008b, hereafter PHHF08).  The parent
sample used by PHHF08 to search for PDLAs excluded broad absorption
line (BAL) QSOs and required a S/N$>$4 over at least one region of 20
consecutive pixels.  The resulting PDLA sample contains 108 absorbers
with \dv\ $<$ 3000 \kms\ and log \nhi\ $>$ 20.3.  In addition to its
size, a second motivation for the adoption of the PHHF08 sample is
that PHHF08 have re-computed the systemic redshifts of those QSOs with
proximate absorbers, accounting for the well-known systematic
blueshifts that are incurred when using rest frame UV lines such as
\lya\ and CIV (e.g. Gaskell 1982).  Throughout this paper, we use the
values of $z_{\rm abs}$ (determined from the centroid of the strongest
low ion) and $z_{\rm em}$ reported by PHHF08 to compute the values of
\dv.  The revised values of $z_{\rm em}$ take into account systematic
shifts between the observed emission lines and those that are
generally considered to be reliable estimators of the systemic value
(such as Balmer and oxygen recombination lines).  Uncertainties on
these systematic shifts are typically a few hundred \kms.  In order to
explore the impact of QSO proximity on DLAs, we extend the nominal
\nhi\ and \dv\ criteria to include proximate absorbers with log \nhi\
$\ge$ 20 and \dv\ $<$ 10,000 \kms.  These cuts are motivated by the
practical limitations of SDSS's resolution and the availability of
improved redshifts in PHHF08.  There are 326 absorbers that fulfill
the \nhi\ and \dv\ criteria.  The sample is reduced by requiring that
the spectroscopic S/N ratio in the $i$-band (as reported in the SDSS
fits header) exceeds 10.  We also exclude one absorber towards a BAL
QSO that was not flagged in the original filter by PHHF08
(J014049.18-083942.5).  The final sample consists of 85 proximiate
absorbers with log \nhi\ $\ge$ 20, \dv\ $<$ 10,000 \kms, S/N $>$ 10
and improved estimates of $z_{\rm em}$ (Table \ref{PDLA_tab}, Figure
\ref{dla_fig}).  Figure \ref{data_histo} shows the distribution of QSO
and absorber properties. 11 absorbers in the final sample have log
\nhi\ $<$ 20.3.  Although DLAs are more strictly absorbers with log
\nhi\ $\ge$ 20.3, and `proximate' absorbers usually considered to be
within $\sim$ 3000 \kms\ of the QSO, for convenience, we will use the
term `PDLA' to refer to the 85 members of the sample studied in this
paper.

\subsection{Control sample}\label{control_sec}

A handful of the strongest metal transitions can be seen in individual
SDSS spectra.  In order to compare the equivalent width (EW)
distributions of the SiII $\lambda$ 1526 line (which is not only
strong but often located redwards of the \lya\ forest) in the PDLAs
and intervening DLA population, a sample of 85 intervening DLAs is
assembled as a control sample.  The control sample is drawn from the
DR5 DLA catalogue of Prochaska \& Wolfe (2009), supplemented with a
non-statistical sample of lower column density absorbers (recall that
the PDLA sample extends to log \nhi\ = 20).  For each PDLA, the
intervening DLA which is the closest simultaneous match in $z_{\rm abs}$,
\nhi\ and QSO fibre $i$ magnitude is determined.  The fibre magnitude
criterion is included in the matching process so that a similar
spectral S/N is achieved.  Once a match is made, the intervening DLA
is not replaced and can not be considered as a match for a different
PDLA.  We also require that the SiII $\lambda$ 1526 line is redwards
of the QSO's \lya\ line, i.e. $z_{\rm em} < 1.255 \times (1+z_{\rm
abs}) - 1$.  A Kolmogorov-Smirnov test (KS) is performed to ensure
that the control sample properties are well matched to the PDLAs.  The
KS probabilities are 92\% ($z_{\rm abs}$), 83\% (\nhi) and 98\% ($i$
magnitude) indicating that there is no difference in the distribution
of these properties between the DLAs and PDLAs.

A second control sample is constructed in order to assess the
extinction properties of the PDLAs via the comparison on the spectral
slopes of QSOs with and without PDLAs (Section \ref{dust_sec}). The
control sample is drawn from the list of DR5 QSOs with $z>2$, no PDLA
and no strong BAL features, as determined from visual inspection.  We
do not exclude QSOs with intervening DLAs because the PDLA sample will
have the same statistical occurence of intervening absorbers.  In
order to reduce the statistical uncertainty in the properties of our
control sample, multiple control spectra are matched to each PDLA
spectrum.  The best match for each QSO with a PDLA in absolute
Petrosian $i$ magnitude and $z_{\rm em}$ is taken from the list of
non-PDLA QSOs (without replacement) resulting in a sample of 85
control QSOs.  The KS probabilities that the $i$ magnitudes and
$z_{\rm em}$ of the PDLA and control QSOs are drawn from the same
distribution are calculated.  The process is repeated, with each
iteration taking the next best simultaneous match in $i$ and $z_{\rm
em}$ until the KS probability of one of the two quantities drops below
30\%.  This process yields 20 control QSOs for every PDLA QSO.

\section{Composite spectra}

\begin{center}
\begin{table*}
\begin{tabular}{lccccccccc}
\hline \hline
Sample    & \# Absorbers & SII $\lambda$1253 & OI $\lambda$1302 & SiIV $\lambda$1393 &  SiII $\lambda$1526 & CIV $\lambda$1548  &   FeII $\lambda$1608 & Z  & $<z_{\rm abs}>$ \\
 & &  (m\AA)  &  (m\AA) &  (m\AA)  & (m\AA) & (m\AA) & (m\AA) & (Z$_{\odot}$)& \\
\hline \hline
All & 85 & 37$\pm$3 & 368$\pm$8 & 242$\pm$6 & 366$\pm$8 & 357$\pm$8 & 180$\pm$6 & 1/34 & 3.18\\ \hline
\dv\ $<$ 3000 \kms & 29  & 45$\pm$5 & 461$\pm$15 & 375$\pm$5 & 455$\pm$5 & 710$\pm$7 & 293$\pm$11 & 1/25 & 3.25 \\
3000 $<$ \dv\ $<$ 6000 \kms & 27  & 36$\pm$4 & 385$\pm$10 & 220$\pm$10 & 345$\pm$7 & 308$\pm$5 & 171$\pm$7 & 1/37 & 3.23  \\
\dv\ $>$ 6000 \kms & 29  & ... & 332$\pm$10 & 203$\pm$5 & 313$\pm$6 & 217$\pm$7 & 145$\pm$5 & 1/43 & 3.07 \\ \hline

\dv\ $<$ 3000 \kms, \nhi$<$20.7 & 18 & 49$\pm$10 & 395$\pm$15 & 358$\pm$8  & 344$\pm$12 & 641$\pm$10 & 180$\pm$10 & 1/37 & 3.30\\
3000 $<$ \dv\ $<$ 6000 \kms, \nhi$<$20.7 & 23 & 30$\pm$9 & 370$\pm$13 & 230$\pm$15 & 372$\pm$10 & 339$\pm$10 & 156$\pm$15 & 1/34 & 3.21 \\
\dv\ $>$ 6000 \kms, \nhi$<$20.7 &20 & ... & 320$\pm$15 & 239$\pm$13 & 326$\pm$15 &244$\pm$10 &121$\pm$10 & 1/40 & 3.16 \\ \hline

\dv\ $<$ 3000 \kms, \nhi$\ge$20.7 &11 &51$\pm$6 & 800$\pm$26 & 430$\pm$13 &787$\pm$12 &750$\pm$20 &383$\pm$15 & 1/12 & 3.15\\
3000 $<$ \dv\ $<$ 6000 \kms, \nhi$\ge$20.7 & 4 & 41$\pm$7 & 441$\pm$15 & 102$\pm$20 & 297$\pm$20 & 315$\pm$25 & 225$\pm$20 & 1/48 & 3.36\\
\dv\ $>$ 6000 \kms, \nhi$\ge$20.7 &9 &... & 295$\pm$10 & 120$\pm$14 &304$\pm$6 &164$\pm$10 &177$\pm$7 & 1/45 & 2.86\\ \hline

\dv\ $<$ 3000 \kms, $L_{1500}<4\times10^{43}$ & 14 & 60$\pm$10 & 532$\pm$17  & 415$\pm$21  & 500$\pm$15 & 735$\pm$20 & 325$\pm$23 & 1/22 & 2.99\\
\dv\ $<$ 3000 \kms, $L_{1500}>4\times10^{43}$ & 15 &$<$120  & 455$\pm$20 & 306$\pm$13 & 355$\pm$15 & 695$\pm$15  & 242$\pm$20 & 1/36 & 3.49 \\
\hline  \hline
\end{tabular}
\caption{\label{EW_tab}Rest-frame equivalent widths for the full
composite sample and three \dv\ sub-samples.  The penultimate column
indicates the metallicity as a fraction of the solar value, according
to the EW(SiII $\lambda$1526)-metallicity relation of Prochaska et
al. (2008a).  The top line refers to the composite shown in Figure
\ref{full_stack}.  The second section refers to the 3 \dv\ composites
shown in Figure \ref{dv_split}.  The third section refers to the low
and high \dv\ samples with low \nhi\ shown in Figure
\ref{dv_lohi_split}.  The fourth section refers to the low and high
\dv\ samples with high \nhi\ shown in Figure \ref{dv_hihi_split}.
The fifth section refers to cuts in QSO luminosity shown in Figure
\ref{lodv_fl_split}.}
\end{table*}
\end{center}

\begin{figure*}
\centerline{\rotatebox{0}{\resizebox{16cm}{!}
{\includegraphics{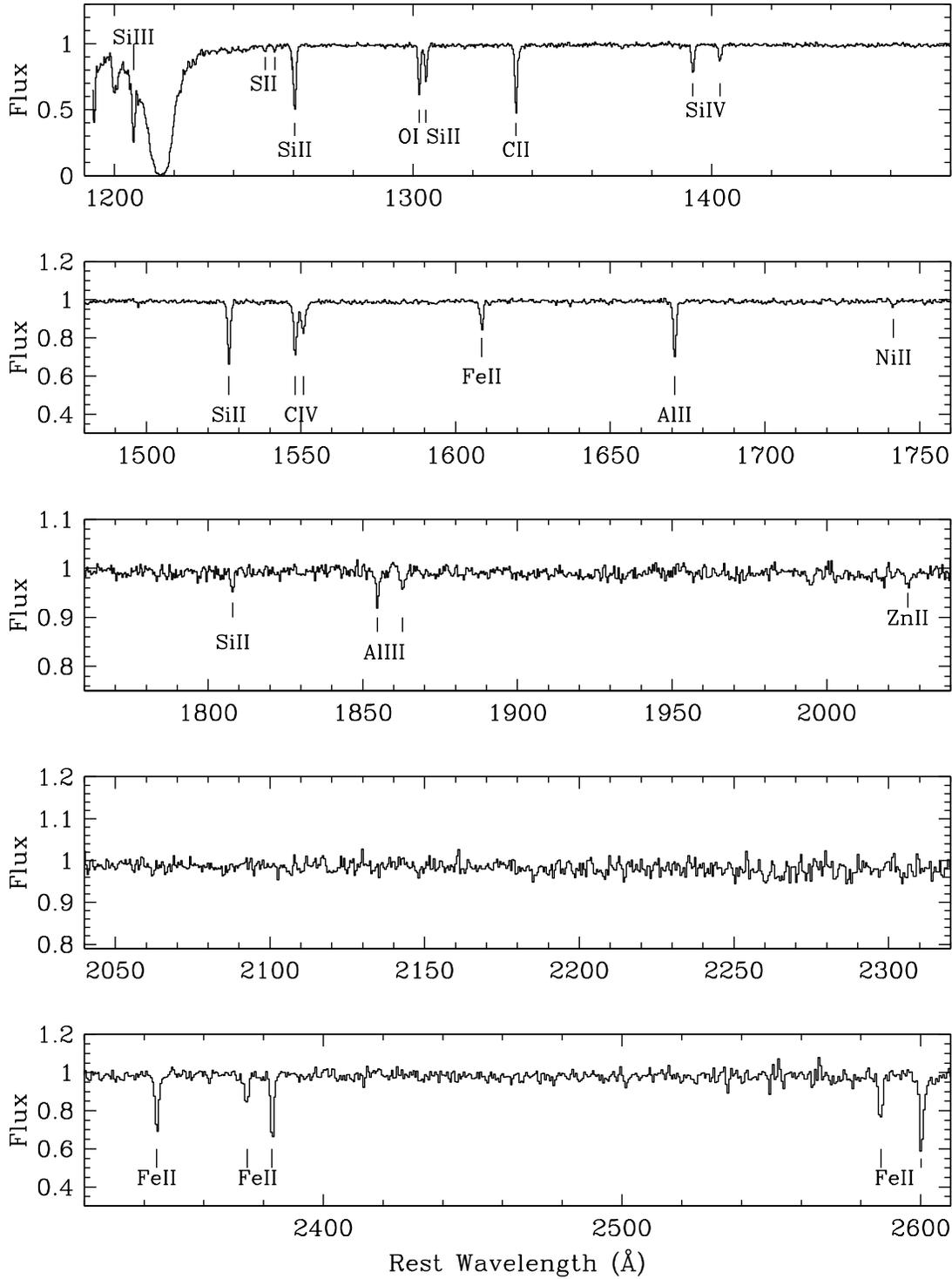}}}}
\caption{\label{full_stack} Median-combined composite spectrum of
85 proximate absorbers with \nhi\ $\ge$ 20 and \dv\ $<$10,000 \kms.  The
$x$-axis is in the absorber rest-frame.  Note the different $y$-axis
scales in the different panels to accentuate absorption feature
detections.}
\end{figure*}

Before creating data stacks, the absorption redshift determined by
PHHF08 was checked by fitting gaussians to three transitions in the
SDSS spectra, where detected: OI $\lambda$1302, SiII $\lambda$1304 and
SiII $\lambda$1526.  The average of the metal line redshifts was
adopted for spectral stacking.  The mean offset between the values
reported by PHHF08 and our mean metal line redshifts is
0.0002$\pm$0.001, where 0.001 corresponds to 75 \kms\ at z=3.  The
SDSS pixel scale is 69 \kms\ per pixel, so the typical shifts are
$\ll$ 1 pixel.  As stated above, the \dv\ values are still calculated
based on the PHHF08 redshifts for easy reference and comparison to
that work.

All of the spectra are shifted to the rest-frame of the proximate
absorber as determined from the metal lines or using the original
\lya\ redshift in 3 cases where no metals were detected.  The spectra
are normalized using continua that are fit using the Starlink software
\textsc{DIPSO}.  Composites are made by median combining the
individual normalized, rest-frame spectra.  Using the median has three
advantages over a mean combination.  First, the median is more robust
against spurious spectral features and a small number of absorbers
with extreme absorption properties.  Second, the unabsorbed continuum
has a value very close to unity in the median composite.  The mean
composite is affected by other absorption (at different redshifts)
resulting in a continuum level noticably below unity.  Nonetheless, we
have performed the analyses in this paper for both mean and median
composites and the basic conclusions are unchanged.  Figure
\ref{full_stack} shows the stack of all 85 absorbers.  The composite
has a S/N $\sim$~150 at $<$2000 \AA, decreasing to $\sim$60 at 2500
\AA.  Note the different $y$-axis scales in the different panels to
accentuate absorption feature detections.  The S/N is high enough to
detect relatively weak species such as SiII $\lambda$ 1808, NiII
$\lambda$ 1741 and ZnII $\lambda$ 2026.

\begin{figure*}
\centerline{\rotatebox{270}{\resizebox{15cm}{!}
{\includegraphics{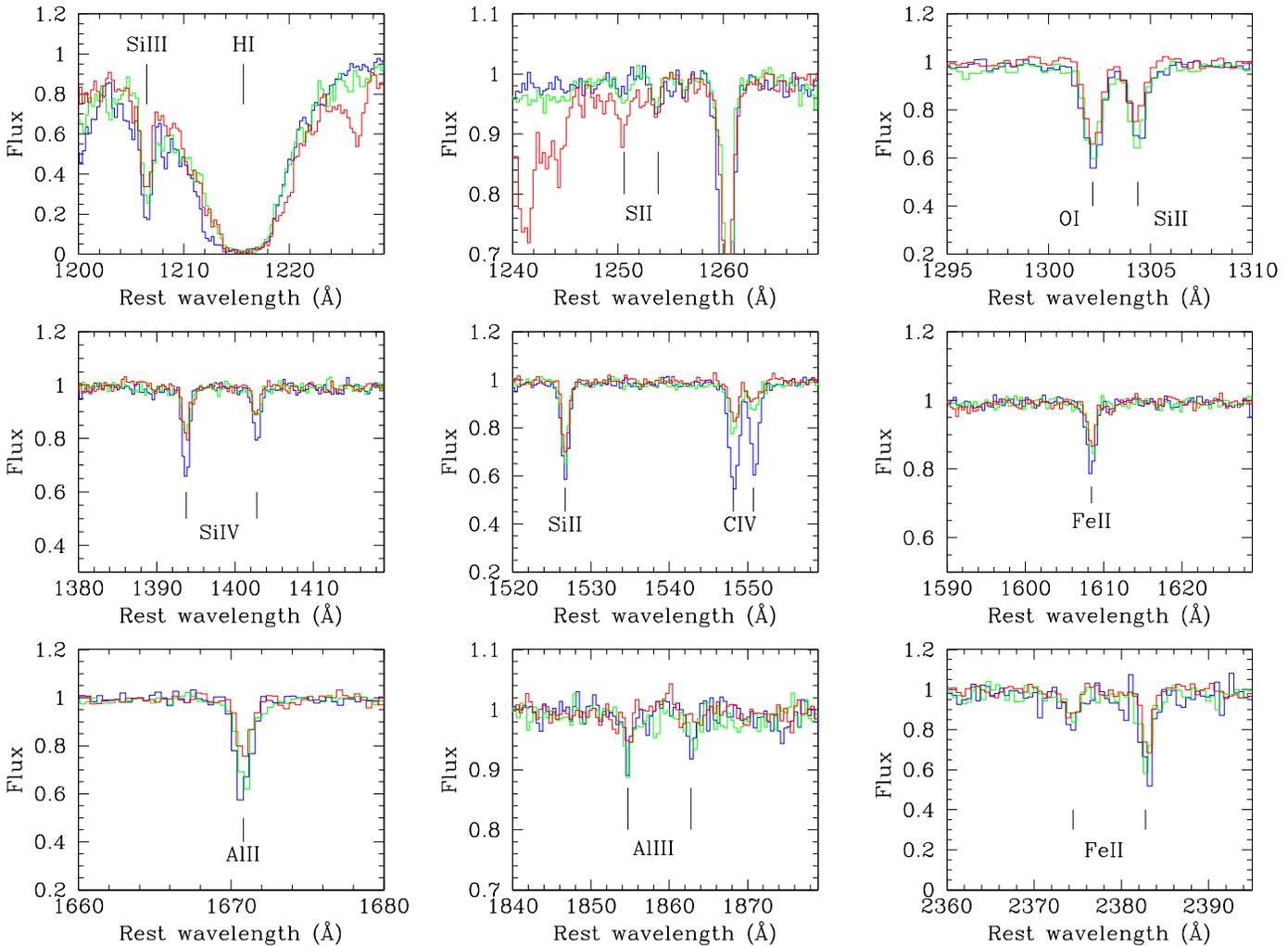}}}}
\caption{\label{dv_split} Median-combined composite spectrum of
proximate absorbers \nhi\ $\ge$ 20 split by \dv: \dv\ $<$ 3000 \kms\ (blue),
3000 $<$ \dv\ $<$ 6000 \kms\ (green), \dv\ $>$ 6000 (red) \kms. The panels
show the strongest absorption features visible in the full composite
shown in Figure \ref{full_stack}. }
\end{figure*}

In Figure \ref{dv_split} we divide the sample into three approximately
equal-sized samples based on \dv.  The cuts are made at \dv\ $<$ 3000
\kms\ (29 absorbers), 3000 $<$ \dv\ $<$ 6000 (27 absorbers) and 6000
$<$ \dv\ $<$ 10,000 \kms\ (29 absorbers).  The median \nhi\ of the
absorbers in each of the velocity sub-samples is 20.6, 20.4 and 20.5
respectively.  PHHF08 quote typical \nhi\ errors of 0.15--0.20 dex,
indicating that the \dv\ sub-sample differences are not significant.  We
also compare each of the samples with a Kolmogorov-Smirnov test and
confirm that there is no statistical distinction in the \nhi\
distributions.  The velocity sub-samples are further divided by \nhi.
When dividing by \nhi\ we do not aim to equalize the number of spectra
contributing to each composite.  Instead, we are motivated by the
physical distinction that might arise as the ISM of the proximiate
absorbers becomes more shielded.
In Figures \ref{dv_lohi_split} and \ref{dv_hihi_split} we show the
composite spectra of QSOs with log \nhi\ $<$ 20.7 and log \nhi\ $\ge$
20.7 absorbers respectively [for clarity, only the lowest (\dv\ $<$
3000 \kms) and highest (\dv\ $>$ 6000 \kms) velocity sub-samples are
shown].  

Assuming that the \nhi\ distributions of the three \dv\ samples are
comparable, for a given metallicity the EWs of the various metal
species should also be comparable in the absence of any proximity
effects.  From Figure \ref{dv_split} it is visually striking that the
majority of absorption lines are strongest in the lowest \dv\
composite.  This is investigated quantitatively by measuring
absorption line equivalent widths.  EWs are determined using single
gaussian fits within \textsc{IRAF}'s \textsc{splot} task. For the CIV
$\lambda \lambda$1548, 1550 doublet the deblending option was used
since the lines have mild overlap in stacks with the highest CIV EWs.
Deblending was also necessary for the OI $\lambda$ 1302 transition in
some of the composites (where blending with SiII $\lambda$ 1304 was
evident).  The rest-frame EW of several detected transitions are given
in Table \ref{EW_tab} and Figure \ref{summary} provides a further
visual representation of how the EWs vary as a function of \dv.  In
most cases, errors in continuum placement dominate over statistical
error.  The EW errors listed in Table \ref{EW_tab} reflect the range
of values obtained over multiple fit realizations rather than the
formal error from the random noise.

\section{Results from composite spectra}

\subsection{Metallicity indicators}

\begin{figure}
\centerline{\rotatebox{0}{\resizebox{9cm}{!}
{\includegraphics{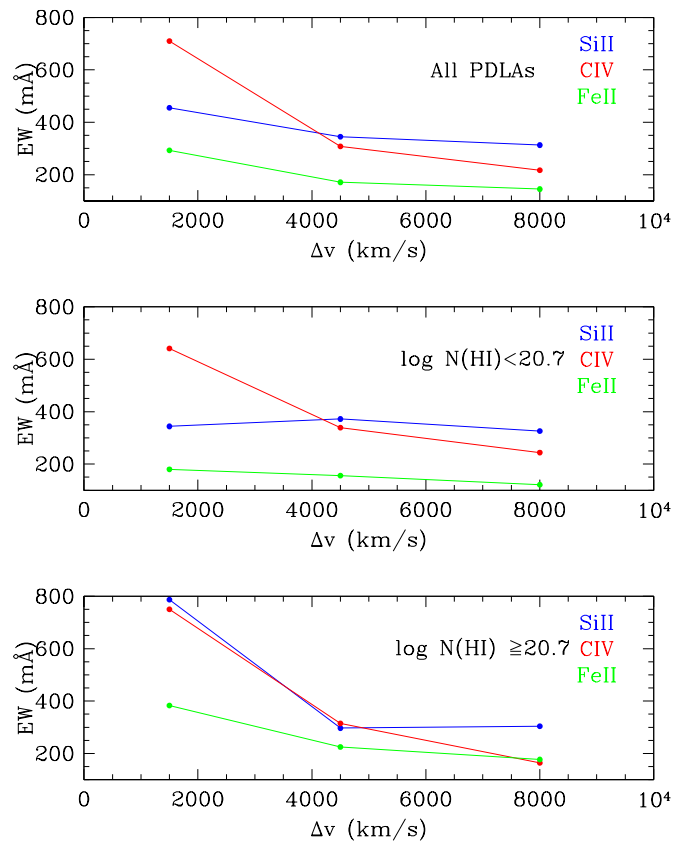}}}}
\caption{\label{summary} For three different absorption species, the
rest-frame equivalent width is plotted as a function of \dv.  The top
panel shows measurements from the composite spectra of the full PDLA
sample (see also the second section in Table \ref{EW_tab}).  The
middle and lower panels show the EWs measured in the composite of spectra of
the low and high \nhi\ sub-samples respectively (third and
fourth sections fo Table \ref{EW_tab}). }
\end{figure}

\begin{figure*}
\centerline{\rotatebox{270}{\resizebox{15cm}{!}
{\includegraphics{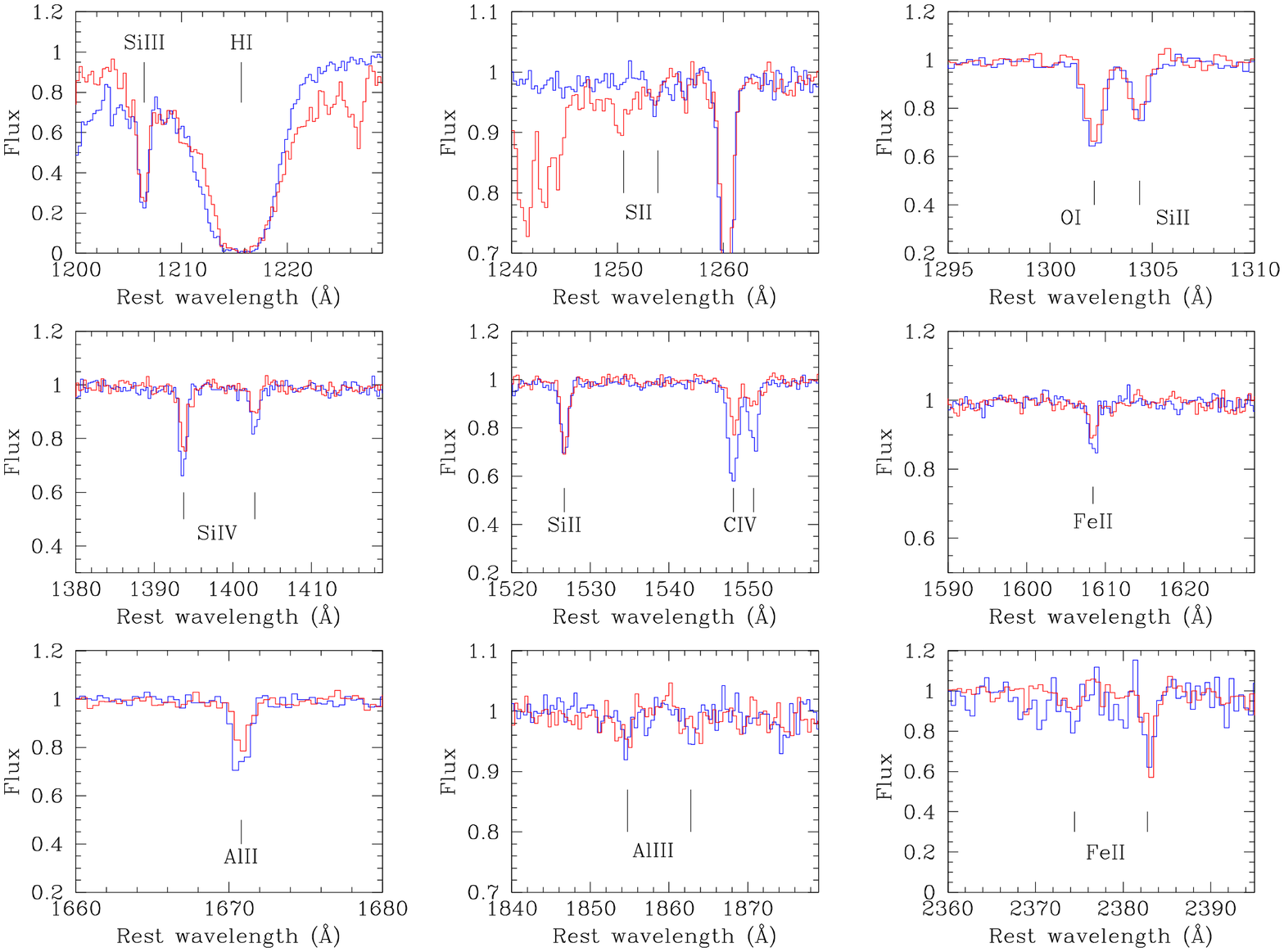}}}}
\caption{\label{dv_lohi_split} Median-combined composite spectrum of
proximate absorbers  with 20 $<$ \nhi\ $<$ 20.7 split by \dv: \dv\ $<$ 3000 \kms\ (blue),
and  \dv\ $>$ 6000 (red) \kms. The panels
show the strongest absorption features visible in the full composite
shown in Figure \ref{full_stack}.  These are the same transitions
shown in Figure \ref{dv_split}. }
\end{figure*}

\begin{figure*}
\centerline{\rotatebox{270}{\resizebox{15cm}{!}
{\includegraphics{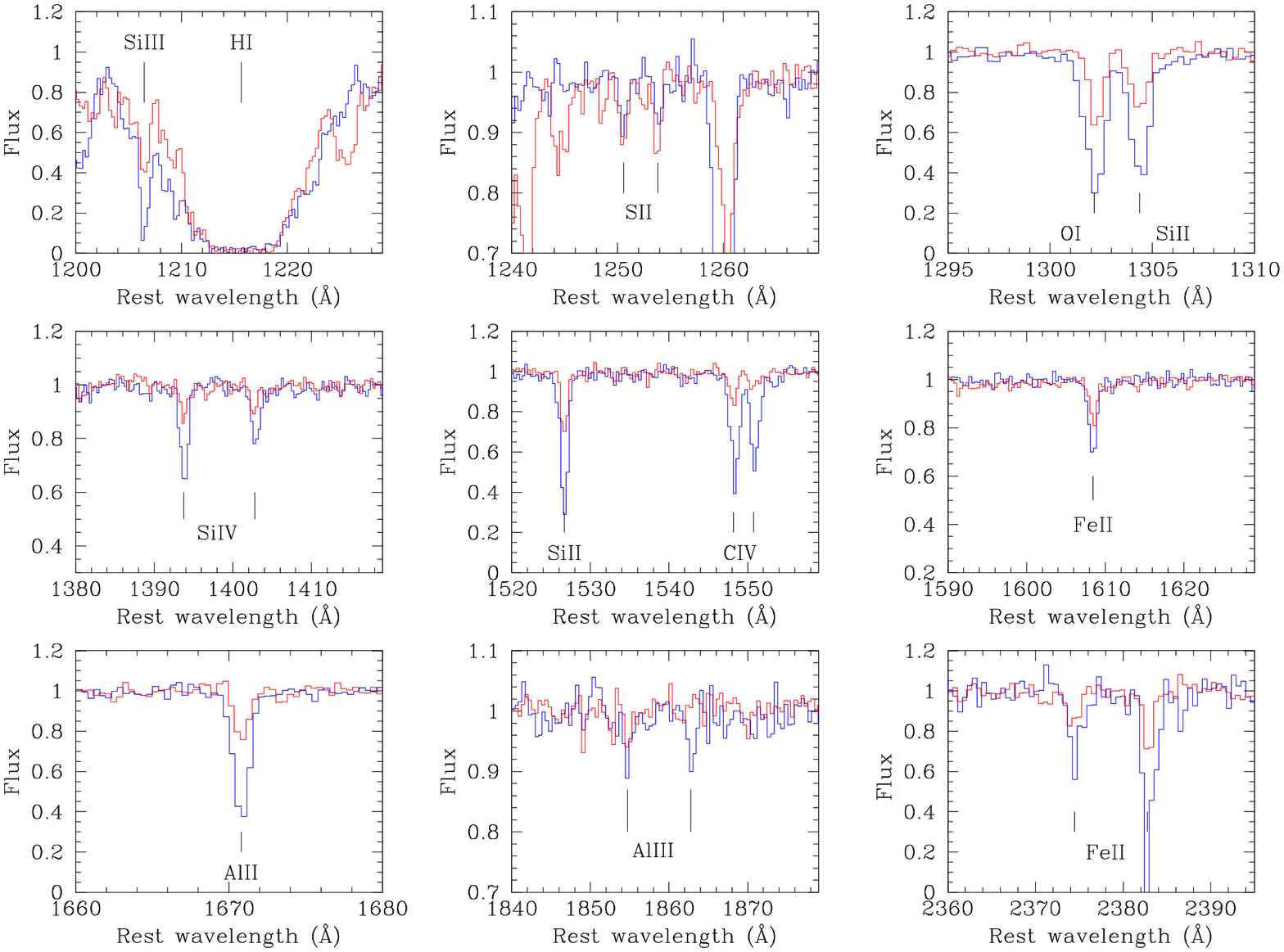}}}}
\caption{\label{dv_hihi_split} Median-combined composite spectrum of
proximate absorbers with \nhi\ $\ge$ 20.7 split by \dv: \dv\ $<$ 3000 \kms\ (blue),
and  \dv\ $>$ 6000 (red) \kms. The panels
show the strongest absorption features visible in the full composite
shown in Figure \ref{full_stack}.  These are the same transitions
shown in Figure \ref{dv_split}. }
\end{figure*}

SiII and FeII are commonly used as indicators of DLA metallicity due
to the relatively large number of observable transitions with a range
of oscillator strengths.  The FeII $\lambda$1608 line is often
unsaturated in intervening DLAs.  For comparable \nhi\ distributions
in the three sub-samples, and under the assumption that the line is
unsaturated, the EW of FeII $\lambda$1608 can therefore be used as a
relative metallicity indicator.  From Table \ref{EW_tab} we see that
the EW of FeII $\lambda$1608 is twice as large in the \dv\ $<$ 3000
\kms\ sample than the \dv\ $>$ 6000 \kms\ composite.  For PDLAs with
3000 $<$ \dv\ $<$ 6000 \kms\ the strength of FeII $\lambda$1608 is
intermediate between the two samples, being $\sim$ 20\% larger than
measured in the high \dv\ composite.  Figure \ref{dv_split} shows that
the same trend is present for other FeII lines.  The mean absorption
redshifts of the composite spectra are given in Table \ref{EW_tab} in
order to indicate any effects that might be attributed to metallicity
evolution.  As a guide, Dessauges-Zavadsky et al. (2009) find that the
[Fe/H] in DLAs changes by 0.19 dex per unit redshift.

Although almost always saturated, the EW of SiII $\lambda$1526 can
also be used as a metallicity diagnostic.  The EWs of heavily
saturated metal lines yield information about the velocity spread of
the absorption (Ellison 2006; Ellison et al. 2008), which in turn is
expected to correlate with mass.  Murphy et al. (2007) and Nestor et
al.  (2003) have both found that there is a correlation between MgII
EW and metallicity, and even unsaturated metal lines appear to exhibit
a correlation between velocity spread and metallicity (e.g. Ledoux et
al. 2006).  Prochaska et al. (2008a) have calibrated the relationship
between EW(SiII $\lambda$1526) and metallicity over two orders of
magnitude from 1/300 to 1/3 solar.  For the values of EW (SiII
$\lambda$1526) in Table \ref{EW_tab} the calibration of Prochaska et
al. (2008a) yields metallicities of 1/25, 1/37 and 1/43 of the solar
value for the low, intermediate and high \dv\ samples respectively.
If no ionization corrections are needed, the FeII and SiII lines
indicate that the metallicity of PDLAs with \dv\ $<$ 3000 \kms\ are
typically twice as high as absorbers with 6000 $<$ \dv\ $<$ 10,000
\kms.

We now consider the metal line strengths within the same velocity
cuts, but now we additionally divide the sample by \hi\ column density
(Figures \ref{dv_lohi_split} and \ref{dv_hihi_split}).  When only the
log \nhi\ $<$ 20.7 absorbers are included in the stack, the EW of FeII
$\lambda$1608 is only 50\% larger in the low \dv\ sample than the high
\dv\ composite (compared to a factor of 2 for the full stack).  The EW
of SiII $\lambda$ 1526 indicates metallicities of 1/37 and 1/40 solar
for the \dv\ $<$ 3000 \kms\ and \dv\ $>$ 6000 \kms\ respectively.  The
only species whose EW increases by more than 50\% in the \dv\ $<$ 3000
\kms\ spectrum is CIV (see middle panel of Figure \ref{summary}).
Conversely, there is a much larger difference in the strengths of
\textit{all} of the metal lines when only absorbers with log \nhi\
$\ge$ 20.7 are included in the composite.  This result is visually
striking in Figure \ref{dv_hihi_split}.  The larger difference between
EWs in the low and high velocity stacks with log \nhi\ $\ge$20.7 is
caused mostly by a large increase (often a factor of two) in the EWs
of transitions in the former.  This is true of both low ionization
species such as SiII $\lambda$ 1526 and high ionization species such
as CIV $\lambda$ 1548.  The EWs of both of these elements are known to
scale with metallicity (Fox et al. 2007a; Prochaska et al. 2008a).
Using the calibration of Prochaska et al.  (2008a), the SiII $\lambda$
1526 EWs indicate metallicities of 1/12 and 1/45 solar for the low and
high \dv\ (\nhi\ $\ge$ 20.7) composites respectively, even though the
former has a slightly higher mean redshift (3.15 compared to 2.86, see
Table \ref{EW_tab}).  This is consistent with the finding of Ellison
et al. (2010) that the PDLAs have a higher metallicity than the
intervening population when the \nhi\ is large.  However, the
difference between EWs in the low and high velocity stacks with log
\nhi\ $\ge$20.7 (Figure \ref{dv_hihi_split}) is exagerated in the case
of CIV and SiIV, whose EWs in the high \dv\ composites significantly
\textit{decrease} at higher \nhi.  The lower column densities of CIV
and SiIV in the higher \nhi\ composite (for \dv\ $>$ 6000 \kms) is
surprising.  At these relative velocities we would consider these
systems to be largely intervening and Fox et al. (2007a) found that
there is no trend between N(CIV) and \nhi\ in intervening DLAs.
However, at CIV column densities above $\sim$ 14.5 the doublet becomes
saturated and only a lower limit can be derived, a fairly common
situation for DLAs (Fox et al. 2007a).  Since Fox et al. (2007a) show
that both the column density and velocity spread of CIV correlate with
metallicity, the lower EW(CIV) at high \nhi\ and high \dv\ may be
another manifestation of the paucity (at intervening redshifts) of
high metallicity absorbers with high \nhi\ (e.g. Schaye 2001; Krumholz
et al. 2009).

\subsection{Ionization indicators}

Ellison et al. (2010) suggested that sub-solar SII/SiII ratios may be
an indicator of a hard ionizing source.  SII is usually studied via
the triplet at 1250, 1253 and 1259 \AA.  In the SDSS data, the
strongest line ($\lambda$ 1259) is blended with the SiII $\lambda$
1260 line.  The next strongest line ($\lambda$ 1253) is clearly
detected in our 3 \dv\ composites, but there may be some contribution
from \lya\ absorption in the \dv\ $>$ 6000 \kms\ sample.  We therefore
only consider the SII $\lambda$ 1253 absorption in the low and
intermediate velocity samples.  Table \ref{EW_tab} shows that although
the low \dv\ composite has a slightly higher (by 9 m\AA) EW, the
measurements are consistent within the errors.  If the silicon
abundance is $\sim$ 50\% higher in the low \dv\ sample, the sulphur
EWs indicate that the \dv\ $<$ 3000 \kms\ PDLAs have a SII/SiII ratio
50\% lower than the 3000 $<$ \dv\ $<$ 6000 \kms\ sample.  Similarly,
the SiII EW in the \nhi\ $\ge$ 20.7 composite (third section of Table
\ref{EW_tab}) has increased by a factor of 2.5 at low \dv, whereas the SII
EW has increased by only 20\% (and is actually
consistent with the 3000 $<$ \dv\ $<$ 6000 \kms\ sample within the errors).
Higher SII/SiII ratios in PDLAs are consistent with the trend
identified by Ellison et al. (2010).  However, the relatively large
uncertainties on the SII EWs mean that this result should be regarded
as tentative.

The species which exhibits the most striking difference between the 3
\dv\ composites is CIV.  There is more than a factor of three
difference between the EW of CIV $\lambda$ 1548 in the low and high
\dv\ composite spectra.  This factor is even larger when only
the log \nhi\ $\ge$ 20.7 are included in the stack, where the
EW in the low \dv\ spectrum is almost a factor of 6 larger than
the \dv\ $>$ 6000 \kms\ composite.  The difference between the
low and high \dv\ samples is much smaller for the low \nhi\
absorbers, only a factor of $\sim$ 2.5.  We see a similar behaviour
for the SiIV EWs: a factor of $\sim$ 3.5 difference between low
and high \dv\ composites when log \nhi\ $\ge$ 20.7, but only
50\% difference for the low \nhi\ stack.  This is ostensibly surprising,
as one might expect the lower \nhi\ absorbers to be more susceptible
to ionization effects, whereas high \nhi\ absorbers might remain
shielded even if they are relatively close to the QSO.


NV is observed in the low \dv\ composite but unfortunately can not be
compared to the intermediate or high velocity spectra, because of
blending by the \lya\ forest.  However, Ellison et al. (2010) have
argued that strong NV in PDLAs may be at least partly due to higher
metallicity, rather than (entirely) due to enhanced ionization (see
also Fox et al.  2009).

\subsection{Dependence on QSO luminosity}\label{luminosity_sec}

\begin{figure*}
\centerline{\rotatebox{270}{\resizebox{15cm}{!}
{\includegraphics{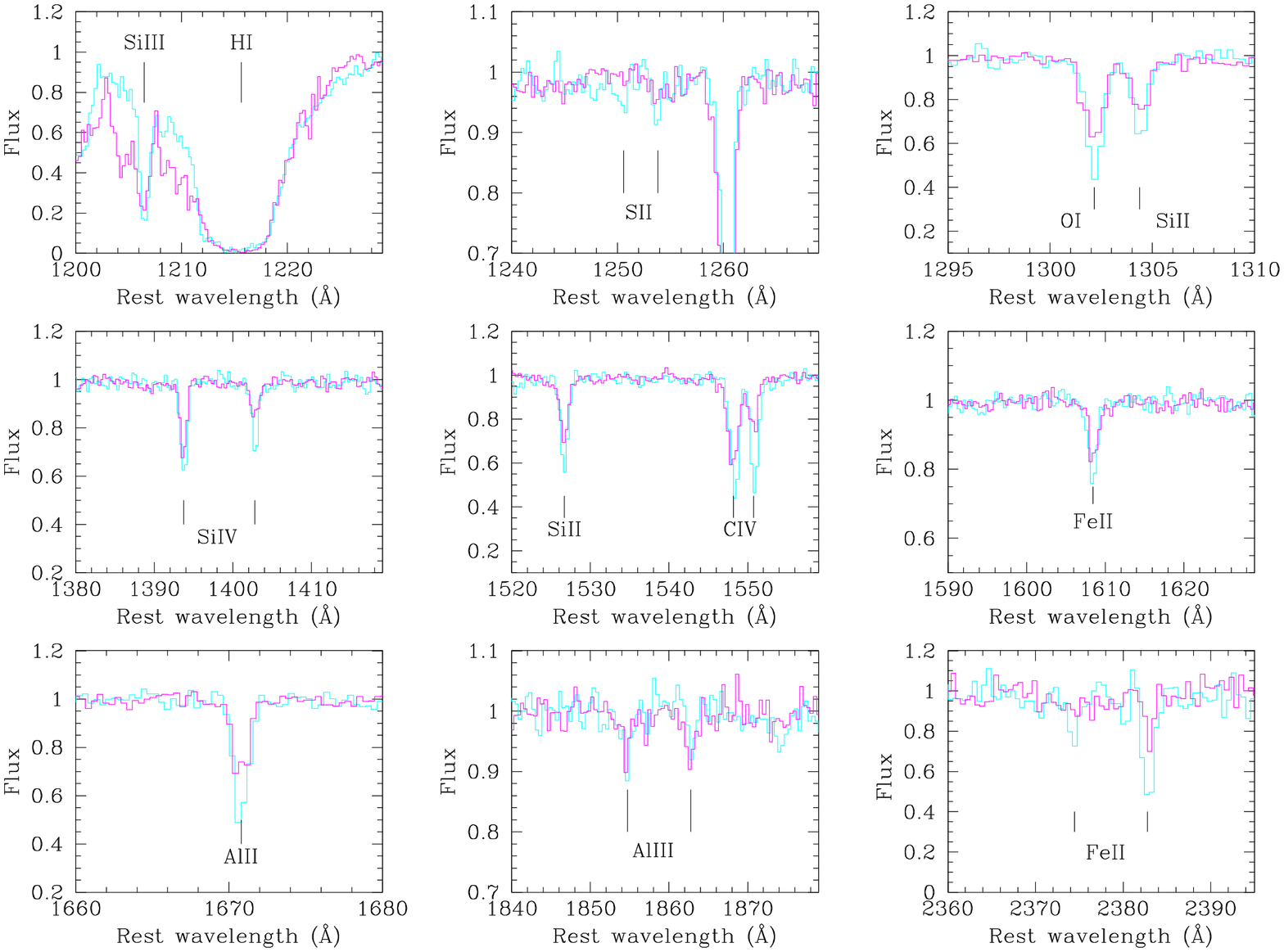}}}}
\caption{\label{lodv_fl_split} Median-combined composite spectrum of
proximate absorbers with \dv: $<$ 3000 \kms, split by QSO luminosity
at $\lambda=1500$ \AA.  The half with the highest luminosities
($L_{1500} > 4\times10^{43}$ ergs/s/\AA) are plotted in magenta and
the half with the lowest luminosities ($L_{1500} < 4\times10^{43}$
ergs/s/\AA) are plotted in cyan. The panels show the strongest
absorption features visible in the full composite shown in Figure
\ref{full_stack}.  These are the same transitions shown in Figure
\ref{dv_split}. }
\end{figure*}

We investigate the effect of QSO luminosity on the composite spectra
by dividing the sample of \dv\ $<$ 3000 \kms\ absorbers (of which
there are 29) in half, based on the QSO flux at $\lambda$=1500 \AA\ in
the QSO rest-frame.  The median luminosity used to divide the QSOs
into `low' and `high' luminosity samples is $4 \times 10^{43}$
ergs/s/\AA.  The profiles of prominent absorption lines are shown for
the low and high luminosity composites in Figure \ref{lodv_fl_split}.
The EWs of selected transitions are given in Table \ref{EW_tab}.  The
EWs of the low and high ionization species alike are larger in the
lower QSO luminosity composite, despite an excellent agreement in the
median \nhi\ values of the two samples (20.6 and 20.5) and a KS test
probability of 86\% that they are drawn from the same distribution.
The difference in the profiles in Figure \ref{lodv_fl_split} is
therefore not due to different \nhi\ distributions and is also
unlikely to be due to an ionization effect for several reasons.
First, OI is expected to be robust against ionization effects (e.g.
Vladilo et al. 2001).  Second, in the presence of either a hard (QSO)
or a soft (stellar) ionizing spectrum, metallicities derived from SiII
or FeII are expected to be over-estimated (Howk \& Sembach 1999;
Vladilo et al. 2001; Rix et al. 2007).  Third, Ellison et al. (2010)
show that [S/H] is under-estimated by at least a factor of three in more
than half of the \nhi\ $<$ 20.7 PDLAs in their high resolution study,
yet the SII EWs in the low and high flux composites are consistent
within the errors.

York et al. (2006) and Vanden Berk et al. (2008) conducted a similar
analysis to ours, constructing composite spectra from SDSS data of
MgII absorbers at intervening and proximate velocities, respectively.
When their samples were divided by QSO magnitude (at $i=19.1$ and $M_i
= 26.5$, for the two studies) the faint QSO composite was found to have
stronger metal lines for both intervening and proximate absorbers.  It
was concluded that this effect was likely due to a selection bias
since weaker MgII systems are typically only detected towards brighter
QSOs.  The (mean) composite spectra of the bright and faint samples
were therefore not constructed of absorbers with the same MgII EW
distribution.

Such a bias is not likely to affect our selection of PDLAs.  First,
the identification of a PDLA relies on the broad, saturated \lya\
which is easily detected in spectra with much lower S/N than we
require here (recall that PHHF08 required a S/N$>$ 4, but our
criterion is an $i$ band S/N $>$ 10).  Second, we have checked that
the \nhi\ distributions of the low and high luminosity samples are
consistent.  This can be seen visually in the top left panel of Figure
\ref{lodv_fl_split} and is confirmed with a KS test (described
above). Therefore, although we might expect the overall S/N of our low
luminosity composite to be lower, this does not affect the EW of the
metal lines we measure in our median composite.  Nonetheless, we check
explicitly that the dependence of PDLA EWs on QSO luminosity is
not due to a bias by constructing composites of \dv\ $>$6000 \kms\
absorbers again split by QSO luminosity.  The EWs of the low and high
ionization lines are now largely consistent with each other, as might
be expected for these absorbers, which most studies would consider as
intervening.  However, a quantitative comparison of the low vs. high
luminosities compsites of the \dv $>$ 6000 \kms\ sample is hampered by
a non-negligible difference in the typical \hi\ column densities.
Both the median and mean \nhi\ values differ by $\sim$ 0.3 dex with a
KS probability of only 6\% that they are drawn from the same
population.

It is not clear why the PDLAs should have higher metal line EWs when
they are within 3000 \kms\ of lower luminosity QSOs, but it is
consistent with an inverse relationship between QSO luminosity and
PDLA metallicity.  One caveat is that although the \nhi\ distributions
are consistent between the low and high luminosity sample, the mean
absorption redshift of the high luminosity sample is $\Delta z$ = 0.5
higher than the low luminosity sample.  Dessauges-Zavadsky et
al. (2009) found that [Fe/H] changes by 0.1 dex over a $\Delta z$=0.5,
whereas the logarithmic difference in metallicites inferred from the
SiII EWs is 0.2 dex.  Given the uncertainty of converting SiII EW to
metallicity, it is premature to conclude that there is a definite link
between QSO luminosity and PDLA metallicity.  However, the concept of
interplay between an AGN and nearby galactic ISM is not without
precedent, either observationally or theoretically.  Irwin et
al. (1987) found that the radio jet axis of a nearby active galaxy was
aligned with an HI tail in a close companion.  They suggested that the
companion's HI gas was being ram-pressure stripped by the AGN jet.
Alignment of the QSO outflow with the line-of-sight has been
previously invoked to explain the higher incidence of transverse DLAs,
relative to the proximate line of sight absorbers (Hennawi \&
Prochaska 2007).  In a case study of one transverse absorber located
only $\sim$ 100 kpc from the QSO, Prochaska \& Hennawi (2009) find
extreme kinematics and consider the case for QSO-absorber
interactions.  The effect of HI stripping has been modelled by Fujita
(2008) and confirmed as an effective mechanism for sufficiently high
energy AGN.  Such interactions could prematurely (or temporarily)
shut-down the star formation leading to lower metallicities.  Such a
scenario might also explain why the lower \nhi\ PDLAs have
metallicities as low as the intervening population (e.g. Table
\ref{EW_tab}, Ellison et al. 2010) whereas the high \nhi\ have been
able to retain their gas and accrue their metals.

\subsection{\lya\ emission in the DLA trough}

\begin{figure}
\centerline{\rotatebox{270}{\resizebox{6cm}{!}
{\includegraphics{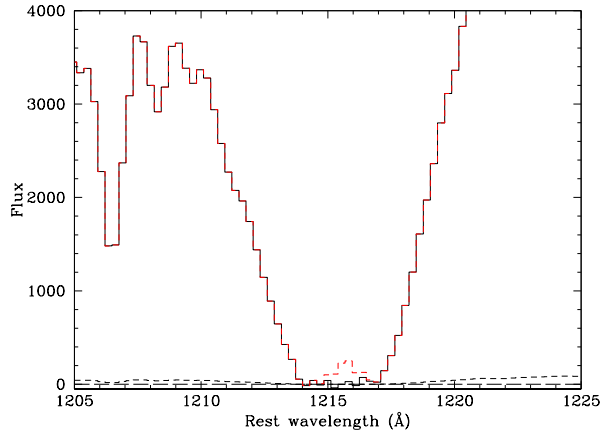}}}}
\caption{\label{lya_emission} A zoom of the \lya\ absorption trough
of the stacked spectrum (black solid line) of the 29 \dv\ $\le$
3000 \kms\ PDLAs.  The red dashed line
shows a simulated stack where each of the 29 PDLAs has an emission line
corresponding to $L_{Ly\alpha}=3 \times 10^{42}$ ergs/s superimposed
at the centre of the \lya\ absorption trough. }
\end{figure}

The detection of \lya\ emission superimposed on the DLA trough is a
rare occurence, but one that is apparently more common in PDLAs than
DLAs (M\o ller \& Warren 1993; M\o ller, Warren \& Fynbo 1998; Leibundgut
\& Robertson 1999; Hennawi et al. 2009).  The luminosities associated
with the \lya\ emission are typically 5--20 $\times 10^{42}$ ergs/s
and may show large velocity widths and spatial extents of tens of kpc
(Fynbo et al. 1999; Hennawi et al. 2009).  

One of the PDLAs in our sample has been recently identified by Hennawi
et al. (2009) as one of the most luminous PDLA \lya\ emitters with a
total \lya\ luminosity of $\sim 4 \times 10^{43}$ ergs/s.  Visual
inspection of the rest of our sample (Figure \ref{dla_fig}) yields no
other candidate \lya\ emitter and there is no emission signal in the
core of the composite spectrum.  In order to determine an approximate
limit for the emission luminosity in our combined spectrum we
artifically superimpose a gaussian emission line in the core of the
PDLA trough in the observed SDSS spectrum for the 29 absorbers within
3000 \kms\ of the QSO.  The same \lya\ luminosity is assumed for every
PDLA, where the total integrated flux then depends only on the
luminosity distance to the absorber.  The width of the emission line
is set to $\sigma$= 100 \kms\ in all cases.  The spectra are then
shifted to the rest-frame with a (1+z)$^4$ scaling to conserve the
total flux and stacked to produce a new composite.  Figure
\ref{lya_emission} shows the stacked spectrum of both the original 29
low \dv\ PDLAs and an overlay of one of our simulated spectra with
emission (in this case where the flux corresponds to a \lya\
luminosity of $L_{Ly\alpha}=3 \times 10^{42}$ ers/s).  Our tests
demonstrate that the composite spectrum is sufficient to detect
luminosities of a few $\times$ 10$^{42}$ ergs/s if such luminosities
are commonplace in PDLAs.  However, this limit is likely to be
optimistic for 2 reasons.  First, in previous detections of \lya\ emission,
the emission line often appears to
be offset by a few hundred \kms\ from the centre of the \lya\
absorption trough.  Stacking analyses are quite sensitive to such
offsets (Ellison et al. 2000).  Second, if the flux is spread out over
many hundreds of \kms, as seen by Hennawi et al. (2009), the broader
profile becomes more difficult to detect.  In the absence of a
well characterized distribution of line offsets and widths, it would
be premature to attempt more sophisticated stacking simulations.

\section{Dust in PDLAs}\label{dust_sec}

Over the past two decades, there has been considerable debate in the
literature regarding the amount of dust extinction associated with
DLAs.  Although these studies may disagree in the details, there is now a
broad concensus in the contemporary literature that the typical dust
reddening is small.  Some examples of recent works on this topic and
detections or limits on the reddening include Murphy \& Liske (2004) [
E(B$-$V) $<$ 0.01], Ellison, Hall \& Lira (2005) [ E(B$-$V) $<$ 0.04],
Vladilo et al. (2008) [ E(B$-$V) $=$ 0.006] and Frank \& Peroux (2010)
who find a best fit E(B$-$V) $= -$0.002, i.e. consistent with no
reddening.

In this section, we investigate whether the PDLAs have extinction
properties that are consistent with the intervening population, since
there are several \textit{a priori} reasons to expect that they may
differ.  As we have demonstrated above, the PDLAs (particularly those
at low \dv\ with high \nhi) have higher metallicities, which appear to
go hand-in-hand with higher gas phase depletion indicators 
(e.g. Prochaska \& Wolfe 2002).
Moreover, Kaplan et al. (2010) have shown that their sample of
metal-strong DLAs (a sample which overlaps considerably with our
PDLAs, as we discuss in the next section) have systematically redder
spectra than a sample of control (non-metal strong) DLAs.  Finally,
proximate MgII absorbers at $z \sim$ 1--2 have reddening values that are
approximately twice that of the intervening MgII systems:  E(B$-$V) =
0.02 and 0.01 respectively (Vanden Berk et al. 2008; York et al. 2006).

There are three main techniques for determining the reddening in a
sample of DLAs.  The first uses broad band colour information, usually
in the optical (e.g. Khare et al. 2004; Vladilo et al. 2006, 2008),
but if IR photometry is also available, a more stringent limit can be
obtained (e.g. Ellison et al. 2005).  If spectral data are available, a
more detailed extinction analysis is possible and there are two main
techniques that have been applied.  The first involves fitting a power
law to unabsorbed regions of the QSO continuum, avoiding the emission
features (e.g. Pei, Fall \& Bechtold 1991;
Murphy \& Liske 2004; Kaplan et al. 2010).  The second
involves making composite spectra of an absorber and non-absorber
sample, where the ratio reveals the average extinction curve of the
former (e.g. York et al. 2006; Wild et al. 2006; Vanden Berk et
al. 2008; Frank \& Peroux 2010).  We have experimented with both
techniques.  The composite method was found to be fundamentally
limited (in our relatively small sample) by our ability to average out
the range of underlying QSO properties.  Although this technique has
been used successfully on smaller samples than ours (e.g. Wild et
al. 2006), the redshift range of our absorbers is such that there are
many emission lines in our rest-frame spectrum.  As noted by Frank \&
Peroux (2010), this leads to residuals and structure in the region
1300-1700 \AA.  Despite various attempts at matching parameters, we
were unable to overcome this limitation and were always dominated by
the underlying variation in emission line properties.  This may be mitigated
in larger samples (e.g. Frank \& Peroux 2010) where residuals lead
to an apparent bluing of only E(B$-$V)$ \sim -0.002$.

\begin{figure}
\centerline{\rotatebox{90}{\resizebox{7cm}{!}
{\includegraphics{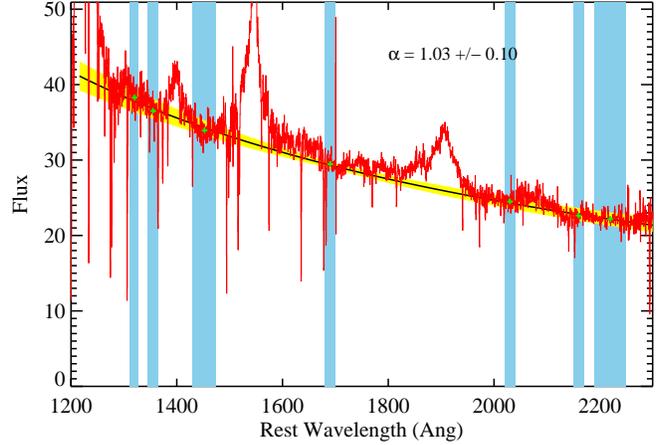}}}}
\caption{\label{index_fit} Example of the spectral fitting procedure.
A power law ($f(\lambda) \propto \lambda^{-\alpha}$) fit is made to the
median value (small green crosses) in each of up to seven continuum
regions (vertical blue shaded bands).  The solid black line shows the
fit and the yellow region shows the uncertainty as a function of
rest-frame wavelength.}
\end{figure}

Instead, we adopted the approach of fitting emission- and
absorption-free regions of the QSO continua with a power law and
examining the difference in spectral indices of the PDLAs and a
control sample of no-PDLA QSOs.  The technique is the same as we have
applied previously in Kaplan et al. (2010).  In brief, the median flux
of up to seven emission-free regions (1312--1328, 1345--1365,
1430--1475, 1680--1700, 2020--2040, 2150--2170, and 2190--2250 \AA) is
calculated from the rest-frame QSO spectrum.  The number of regions
depends on the QSO redshift: Higher redshift QSOs will have few
spectral regions to fit, as the redder regions are shifted beyond the
SDSS spectral coverage.  A power law of the form $f(\lambda) \propto
\lambda^{-\alpha}$ is fit to these median values where reddening due
to dust will lead to smaller values of $\alpha$.  An example of the
fitting procedure is shown in Figure \ref{index_fit}.  Three QSOs have
2 PDLAs in their line of sight.  These were excluded from the
analysis, leaving a sample of 79 PDLAs.  Fitting is also performed for
the 20 control QSOs (Section \ref{control_sec}) matched to each of the
79 PDLAs (i.e. a total of 1580 control spectra).  The use of control
spectra matched in redshift and magnitude will help circumvent any
bias from the dependence of continuum slope on luminosity or the
number of wavelength windows fitted.  We have additionally checked
that there is no overall trend of $\alpha$ with redshift in the
control sample.

\begin{figure}
\centerline{\rotatebox{0}{\resizebox{8cm}{!}
{\includegraphics{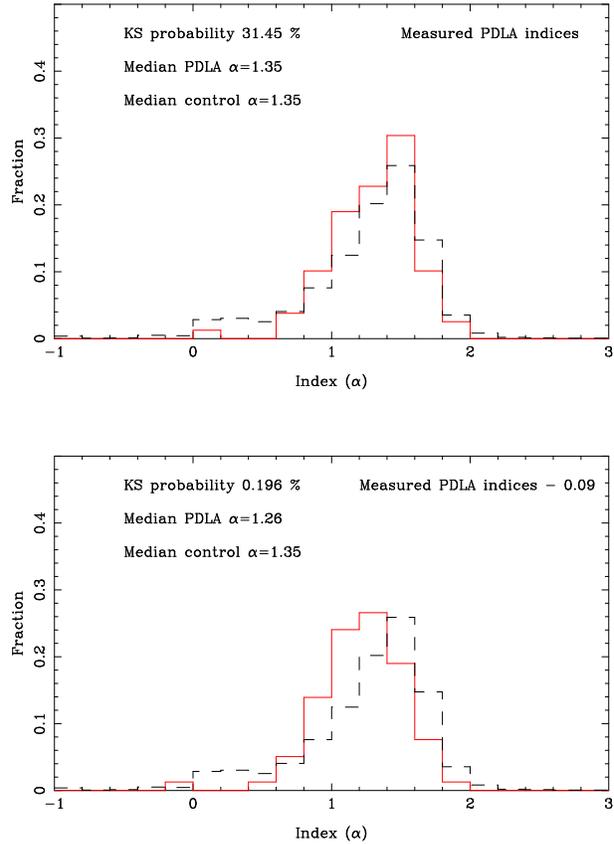}}}}
\caption{\label{alpha_hist} Fractional distribution of power law
indices for the PDLA sample (red solid histogram) and matched
control QSO sample (black dashed histogram).  There are 20 control
QSOs for every QSO with a PDLA.  The upper panel shows the
distribution of measured PDLA indices.  The lower panel shows the
PDLA indices decreased by a value of 0.09 which is the offset required
to yield a KS value less than 0.3 \% (3 $\sigma$ rejection of the null
hypothesis).  The control histrogram
is the same in both panels.}
\end{figure}

The top panel of Figure \ref{alpha_hist} shows the fractional
distribution of power law indices in the PDLA and control sample.  A
KS test indicates that there is no statistical difference between the
distributions and the two samples have the same median value.  In
order to determine a limit for the reddening in our sample, we
estimate the minimum change in the indices of the PDLA sample that
would have resulted in a KS test probability of 0.3\% that it is drawn
from the same distribution as the control (corresponding to a 3
$\sigma$ certainty).  Small incremental increases are made to the
observed indices of the QSOs with PDLAs and the KS probability
(relative to the control) is re-computed until the value reaches
0.3\%.  Considering all 79 PDLAs, decreasing all the indices by a
single value of 0.09 yields a KS probability below 0.3\%.  The new
distribution of PDLA indices is shown in the lower panel of Figure
\ref{alpha_hist}.

The sample of 79 PDLAs includes absorbers with \dv\ values as large as
10,000 \kms.  At such large velocity separations the absorbers are
likely to be dominated by the intervening population.  However,
we find that comparing only the lower \dv\ subsamples with
their control still yields no significant difference in the
distribution of $\alpha$.  There is also no difference in the continuum slopes
when the sample is split by \nhi.  We repeat
the above procedure for determining the minimum difference
in $\alpha$ that would result in a significant KS result.    For 
the \dv\ $<$ 6000 \kms\
absorbers (51 PDLAs) we again find an index decrease of 0.09 before the
KS probability drops below 0.3\%.  For the \dv\ $<$ 3000 \kms\ (28
absorbers) the test is less sensitive and an index change of 0.19 is
required before the distribution is significantly different from the
control.  Using the smaller data set therefore provides the more stringent
constraint on reddening.

To convert the index change that results in a significant KS
probability into a measure of dust extinction, we define the
difference between the control and PDLA indices as $\delta \alpha =
\alpha_c - \alpha_P $.  As described in Kaplan et al.  (2010), we can
write the colour excess at two wavelengths as a function of $\delta
\alpha$:

\begin{equation}
E_{\lambda_1 - \lambda_2} = -2.5 . log_{10}  \left[\left(\frac{\lambda_1}{\lambda_2}\right)^{\delta \alpha}\right] .
\end{equation}

The values of $\lambda_1$ and $\lambda_2$ are set to be 1300 and 2000
\AA\ respectively, as values that are typical of the coverage of our
spectra.  In order to convert this colour excess at $\lambda_1$ and
$\lambda_2$ into the more standard notation of \ebmv, we adopt the SMC
extinction curve of Pei (1992).  We define $\xi(\lambda) = A_{\lambda} / A_V$,
so that\footnote{Pei (1992) defines $\xi(\lambda) = A_{\lambda} / A_B$, so we
have included a factor of 1.33 to the values of $\xi$ derived from the
Pei extinction curve.}

\begin{equation}
E_{\lambda_1 - \lambda_2} = A_{\lambda_1} - A_{\lambda_2} = A_V (\xi_{\lambda_1} - \xi_{\lambda_2}).
\end{equation}

Combining these equations for a $\delta \alpha$ = 0.09 and 0.19 we
derive an $A_V = 0.020$ and 0.041 respectively.  To convert to an
\ebmv\ we use the standard definition $R_V = A_V / E(B-V)$ with $R_V =
2.93$.  For a $\delta \alpha$ = 0.09 and 0.19 we arrive at upper limits
to the colour
excesses of \ebmv\ = 0.007 (\dv $<$ 10,000 \kms\ and \dv $<$ 6000
\kms\ samples) and 0.014 (\dv $<$ 3000 \kms\ sample) respectively.

\begin{figure}
\centerline{\rotatebox{270}{\resizebox{6cm}{!}
{\includegraphics{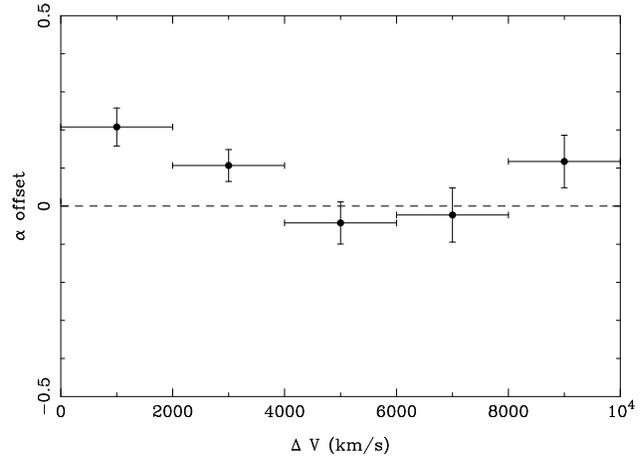}}}}
\caption{\label{alpha_offset} $\alpha$ offset is defined as the index
of the PDLA QSO spectrum minus the mean value of the index of its 20 control QSOs.
The mean $\alpha$ offset and 1$\sigma$ error bars are plotted in bins
of \dv.  Based on a random sampling of the full $\alpha$ offset
distribution, the high value seen in the \dv\ $<$ 2000 \kms\ bin has a
5\% chance of occurring by chance in a sample of this size.}
\end{figure}

An alternative technique for identifying differences in the PDLA and
control sample is to calculate the difference between the $\alpha$ of
the PDLA spectrum and the mean of its 20 control galaxies.  These
`$\alpha$ offsets' are plotted as a function of \dv\ in Figure
\ref{alpha_offset} and should be zero if the QSO with the PDLA has a
continuum slope that is the same as the mean of its controls.
Negative values of the $\alpha$ offset indicate that the PDLAs are
redder than their controls, on average.  Figure \ref{alpha_offset}
appears to show that QSOs with PDLAs whose \dv\ $<$ 2000 \kms\ are
\textit{bluer} than their matched controls.  At larger \dv\ there is
no significant ($> 3 \sigma$) offset.  Even though we take the median
flux in the continuum windows to account for the effect of absorption
features, one may be concerned that indices may be biased in certain
velocity intervals as strong absorption lines move through.  In order
to make the PDLA spectrum bluer, additional absorption at the red end
of the spectrum would be required.  A review of Figure
\ref{full_stack} shows that the only strong lines beyond 1700 \AA\ are
the FeII lines at $\lambda_0 > 2300$ \AA\ and are therefore too red to ever
enter any of our continuum windows.  Intrinsic dependences of the
index on QSO redshift and magnitude are accounted for in the matching
procedure.  Drawing samples at random from the full $\alpha$ offset
distribution indicates that the positive offset seen at \dv\ $<$ 2000
\kms\ would be expected 5\% of the time for bins of this size.  Larger
samples are therefore required to determine whether this result is due
to small number statistics.

\section{Discussion}\label{discussion_sec}

A similar analysis to the one presented here has been performed for
415 proximate MgII absorbers (EW $>$ 0.3 \AA) within \dv\ $<$ 3000
\kms\ of the QSO by Vanden Berk et al. (2008).  They find that the low
ionization lines (MgII and AlII) have similar equivalent widths in the
proximate and intervening composites, whereas the high ionization
lines (CIV and SiIV) are stronger in the \dv\ $<$ 3000 \kms\ sample.
Vanden Berk et al (2008) conclude that this is caused by higher
ionization in the proximate MgII systems.  Although we also find that
the CIV and SiIV EWs are 2--3 larger in the PDLAs at \dv\ $<$ 3000
\kms, compared to the \dv\ $>$ 6000 \kms\ sample, the low ionization
lines such as SiII and FeII are also up to a factor of 2 stronger.
The relative strengths of low and high ionization lines is probably
related to the \nhi\ of the sample.  The MgII absorbers in Vanden Berk
et al. (2008) are likely to have \hi\ column densities that extend
down to log \nhi\ $\sim$ 17.5.  Their composites will therefore be
skewed towards lower \nhi\ values than ours.  When we consider the
lower half of our \nhi\ distribution (log \nhi\ $<$ 20.7) we also find
that the EWs of the low ions become more consistent in the low and
high \dv\ samples (Figures \ref{summary} and \ref{dv_lohi_split}).
The clearest distinction in the metallicities of \dv\ $<$ 3000 \kms\
relative to larger velocity separations is also seen when the \nhi\ is
large (see also Ellison et al. 2010).  This may again explain why
Vanden Berk et al. (2008) did not find evidence for enhanced
metallicities in their proximate MgII sample, whereas our PDLAs show
systematically larger low ionization metal line EWs at small \dv.

\begin{figure}
\centerline{\rotatebox{0}{\resizebox{8cm}{!}
{\includegraphics{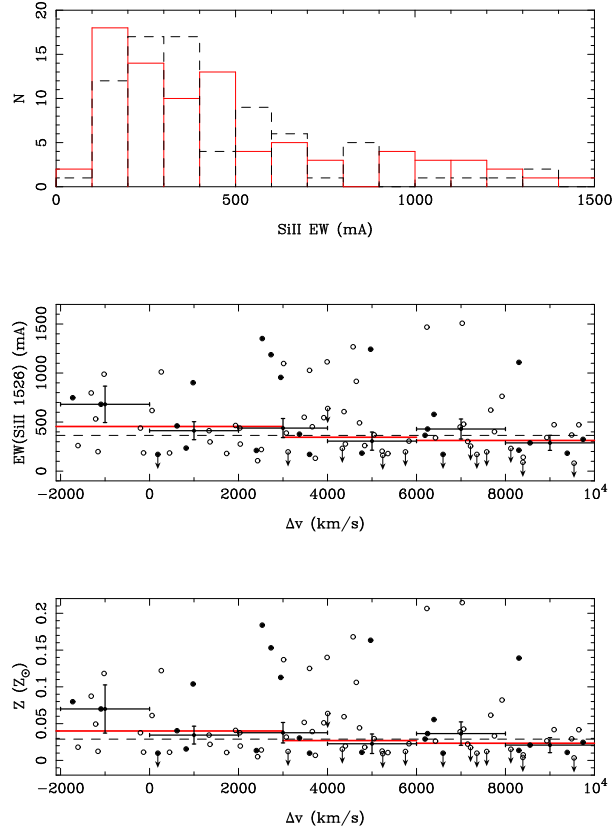}}}}
\caption{\label{bin_SiII} Top panel: Distribution of SiII $\lambda$
1526 EWs for the PDLAs (solid red line ) and matched control sample of
intervening DLAs (dashed black line).  Middle panel: Rest-frame EWs of
the SiII $\lambda$ 1526 line as a function of velocity offset from the
QSO.  Limits are 3$\sigma$.  The open circles show individual PDLA
measurements and the solid circles are median values in velocity bins
of width 2000 \kms. The median is selected so that the small number of
SiII upper limits can be considered as detections without affecting
the binned value.  The horizontal dashed line is the median value of
SiII $\lambda$ 1526 EWs of the intervening control sample.  Lower
panel: SiII $\lambda$ 1526 EWs have been converted to metallicities
(relative to the Sun) using the calibration of Prochaska et
al. (2008a): log $Z/Z_{\odot}$ = 1.41 log(W) $-$ 0.92, where W is the
SiII EW in \AA. The horizontal dashed line is the median metallicity
calculated from the SiII $\lambda$ 1526 EWs of the intervening control
sample.  Errors are 1$\sigma$.}
\end{figure}

Figure \ref{dv_split} and Table \ref{EW_tab} show that the majority of
the transitions in the 3000 $<$ \dv\ $<$ 6000 \kms\ composite exhibit EWs that
are intermediate between the low and high velocity composites.  This
indicates that the 3000 -- 6000 \kms\ absorbers may not yet have
reached the same typical values as intervening absorbers.  We can
attempt to achieve some finer \dv\ resolution by utilizing the SiII
$\lambda$ 1526 line which is detected in 72/85 of the individual SDSS
PDLA spectra.  In cases of a non-detection, the rest-frame 3 $\sigma$ limit
was calculated from the rest-frame spectrum using the equation 
$3\sigma = \frac{3 \times FWHM}{S/N}$.  The FWHM at 1526 \AA\ is
taken to be 0.85 \AA\ (based on the nominal SDSS resolution of
$R=\lambda / \Delta \lambda = 1800$).  The EW of this (usually
saturated) SiII line correlates with metallicity (Prochaska et
al. 2008a) so can be used as an approximate metallicity metric.  In
Figure \ref{bin_SiII} we show the median SiII $\lambda$ 1526 EW in
bins of \dv\ (middle panel) and the corresponding metallicity calibrated
by the Prochaska et al. (2008a) relation (bottom panel).  The enhanced
metallicity is seen most convincingly in the lowest velocity bin (\dv
$<$ 0 \kms) and is due to a combination of a lack of low EWs,
as well as values that range up to 1 \AA\ in equivalent width.  The results
earlier in this paper would further lead us to expect that this will
depend on the \nhi\ of the individual absorbers, but we do not have
the numbers to test both the \dv\ and \nhi\ distributions separately
in this way.

Although the median SiII EW is only enhanced at \dv\ $<$ 0 \kms,
Figure \ref{bin_SiII} shows a notable population of high EW SiII absorbers
at larger velocity separations, particularly around 3000 $<$ \dv\ $<$
5000 \kms.  These are likely to be the absorbers that lead to the
generally larger EWs measured in the intermediate velocity composite
(3000 $<$ \dv\ $<$ 6000 \kms).  Although SiII EW is not a very
accurate estimator of metallicity, the detection of SiII $\lambda$
1808 (with EWs$>$ 150 m\AA) in approximately half of the 3000 $<$ \dv\
$<$ 5000 \kms\ PDLAs with SiII EW $>$ 900 m\AA\ is strongly suggestive
of high metallicities.  It may be surprising that QSO proximity is
responsible for enhanced metallicities at \dv\ $>$ 3000 \kms\ (which
corresponds to a Hubble flow distance of $\sim$ 10 proper Mpc at
$z=3$).  Although the redshifts derived from the SDSS spectra by
PHHF08 are improved from those determined from \lya\ and CIV emission
lines alone, they are likely still only \textit{statistically}
accurate to within a few hundred \kms\ (Shen et al. 2007).  However,
the presence of a population of PDLAs with $-2000 <$ \dv\ $<$ 0
indicates that a combination of non-Hubble flow velocities and
redshift errors can account of $\sim$ 2000 \kms\ in velocity range.
Individual IR spectra of these QSOs are required in order to
determine accurate relative velocities.  The uncertain nature of these
intermediate \dv\ absorbers means that, even with a sample of 85 SDSS
spectra, it is not possible to determine what the correct cut-off
should be for studies that aim to sample the truly intervening
population.  Our results indicate that the largest effect occurs at
velocities well below the cut-offs typically used in DLAs surveys
($\sim$ 3000 \kms).  However, a more conservative approach (until a
larger statistical study can be undertaken) would be to increase the
\dv\ cut-off from 3000 to 6000 \kms.

Although the majority of DLAs are metal-poor, a few rare examples of
high metallicities have been reported in the literature, although
mostly in the lower \nhi\ sub-DLAs (e.g. Prochaska et al. 2006; Peroux
et al. 2008; Meiring et al. 2008; Dessauges-Zavadsky, Ellison \&
Murphy 2009).  The very large samples of QSOs assembled by the SDSS
has also permitted the identification of unusually high metallicity
systems.  Herbert-Fort et al. (2006) identified absorbers with large
metal line equivalent widths from the SDSS DR3 spectra.  These
metal-strong (MS) DLAs are of particular interest for detecting
rare elements, such as boron, germanium and cobalt
(Ellison, Ryan \& Prochaska 2001; Prochaska et al. 2003).
It is
estimated that this metal-strong population represents only
$\sim$ 5\% of the DLA population.  Kaplan et al. (2010) have
determined abundances for a sample of 16 MS DLAs finding a median
metallicity $Z \sim 1/5 Z_{\odot}$ at $z \sim 2$, a factor of around 4
higher than their control sample of non-MS DLAs.  We have compared our
sample of 85 PDLAs (Table \ref{PDLA_tab}) to the list of MS DLAs in
Herbert-Fort et al. (2006).  11 PDLAs (13\%) are present in the
compilation of Herbert-Fort et al. (2006), see Table \ref{MSDLA_tab}.
However, only 45/85 of the QSOs in Table \ref{PDLA_tab} were present
in the DR3 (recall that our sample is drawn from the DR5).  Therefore,
whereas only $\sim$ 5\% of intervening DLAs are metal-strong, the
percentage rises to 11/45=24\% for PDLAs.

\begin{figure}
\centerline{\rotatebox{0}{\resizebox{8cm}{!}
{\includegraphics{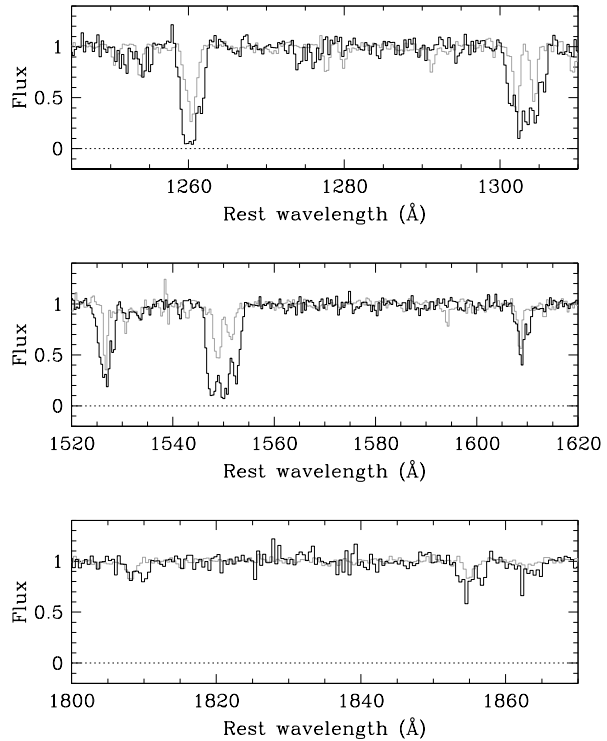}}}}
\caption{\label{plot_J1604} Selected regions of the SDSS spectrum of 
J160413.97+395121.9 (black) with the SDSS spectrum of the
proto-typical metal-strong DLA J081240.68+320808.6 (grey) overlaid.
The top panel includes detections of SII $\lambda \lambda$ 1250, 1259,
SiII $\lambda$ 1260, OI $\lambda$ 1302 and SiII $\lambda$ 1304.
The middle panel includes detections of SiII $\lambda$1526, CIV $\lambda
\lambda$ 1548, 1550 and FeII $\lambda$ 1608.  The lower panel
covers SiII $\lambda$ 1808 and Al III $\lambda \lambda$ 1854, 1862. }
\end{figure}

One of the PDLAs in our sample is the proto-typical MS DLA
J081240.6+320808\footnote{We emphasize again that for convenience, in
this paper we are using the term `PDLA' for absorbers with log \nhi\
$>$ 20 and \dv\ $<$ 10,000 \kms, which is the range over which we look
for trends.  A stricter definition would adopt the standard DLA
criterion of log \nhi\ $\ge$ 20.3.  Moreover, at large \dv\ our sample will
include an increasing fraction of intervening absorbers.}.  In a case
study of this absorber, Prochaska et al.  (2003) used data obtained
with HIRES to make the first detections of several elements never
before detected outside the local group.
J081240.6+320808 is notable because it shows SiII $\lambda$ 1808 (EW
$\sim$ 260 m\AA) absorption even in the SDSS spectrum.  As described
above, SiII $\lambda$ 1808 is observed in several of the individual
SDSS spectra, including J075901+284703, J113008+535419,
J135305$-$025018, J160413.97+395121 and J210025.03-064146 which appear
in the MS DLA catalogue of Herbert-Fort et al. (2006).  To demonstrate
the large EWs of these metal-strong PDLAs, in Figure \ref{plot_J1604}
we show the SDSS spectrum of J160413.97+395121 compared with that of
J081240.6+320808.  J160413.97+395121 has been subsequently studied at
high resolution by Ellison et al. (2010), who confirm a relatively
high metallicity $\sim$ 1/16 $Z_{\odot}$.  J210025.03-064146.0 was
observed with ESI by Herbert-Fort et al. (2006) and measured to have
an even higher metallicity: $Z \sim$ 1/5 $Z_{\odot}$.

\begin{center}
\begin{table}
\begin{tabular}{lcccr}
\hline 
QSO                 & $z_{\rm em}$ &  $z_{\rm abs}$ & log \nhi\ & \dv\ (\kms)\\ \hline
J075901.28+284703.4  & 2.8550 & 2.8223 & 21.05 & 2532.00 \\
J081240.68+320808.6  & 2.7045 & 2.6259 & 21.30 & 6391.40 \\
J095817.81+494618.3  & 2.3555 & 2.2909 & 20.65 & 5831.18 \\
J101725.88+611627.5  & 2.8069 & 2.7681 & 20.60 & 3073.19 \\
J102619.09+613628.8  & 3.8442 & 3.7853 & 20.35 & 3657.30 \\
J113008.19+535419.8  & 3.0495 & 2.9870 & 20.25 & 4575.67 \\
J122040.23+092326.8  & 3.1462 & 3.1322 & 20.75 & 978.39  \\
J135305.18-025018.2  & 2.4149 & 2.3618 & 20.30 & 4647.59 \\
J160413.97+395121.9  & 3.1542 & 3.1625 & 21.75 & -656.45 \\
J210025.03-064146.0  & 3.1295 & 3.0918 & 21.05 & 2729.34 \\
J232115.48+142131.5  & 2.5539 & 2.5729 & 20.60 & -1607.98\\
\hline 
\end{tabular}
\caption{\label{MSDLA_tab} PDLAs identified as `metal-strong'
in the DR3 catalogue of Herbert-Fort et al. (2006). }
\end{table}
\end{center}


Although the PDLAs and MS DLAs may both preferentially sample the high
metallicity end of the DLA distribution, they differ markedly in their
extinction properties.  Whereas the general DLA population causes very
little reddening to the background QSO (Murphy \& Liske 2004; Ellison
et al. 2005; Vladilo et al. 2008; Frank \& Peroux 2010), Kaplan et al
(2010) find significantly flatter power law slopes in QSOs behind
MS DLAs.  Using the methodology described in Section \ref{dust_sec} and
adopting our choice of SMC extinction curve and $R_V$, their report
value of $\delta \alpha = 0.29$ corresponds to $A_V = 0.063$ and
\ebmv\ = 0.022.  In contrast, our most stringent limit (derived for
the \dv\ $<$ 3000 \kms\ sample) gives $A_V <$ 0.041 and \ebmv\ $<$
0.014 for the PDLAs. The reddening by PDLAs is also less than for
associated MgII systems at $z \sim$ 1--2 [E(B$-$V) = 0.02, Vanden Berk
et al. 2008], although this may be at least in part due to the
redshift evolution of reddening in MgII absorbers (Menard et
al. 2008).  However, our detection limit is not sufficiently sensitive
to determine whether or not the PDLAs are more or less dusty than
their intervening cousins.

We have presented evidence in this paper and in Ellison et al. (2010)
that some PDLAs (those with small \dv\ and large \nhi) appear more
metal-rich than the intervening population.  However, the sample of
PDLAs in Ellison et al. (2010) includes an absorber with one of the lowest
metallicities measured to date (J0140$-$0839, [O/H]=$-2.72$).
In Section \ref{luminosity_sec} we have shown that for PDLAs with \dv\
$<$ 3000 \kms, the composite SDSS spectra show weaker metal lines when
the QSO luminosity is higher.  J0140$-$0839 (\dv = 1250 \kms) is
included in the PDLA sample of PHHF08, but excluded by us because the
QSO shows mild BAL features.  Nonetheless it is intriguing that
J0140$-$0839 has a relatively large luminosity: $L_{1500} = 1.3 \times
10^{44}$ ergs/s/\AA, a value exceeded by only two of the QSOs in our
sample.  There are 4 QSOs in our sample with $L_{1500} > 1 \times
10^{44}$ ergs/s/\AA: J082638.59+515233.2, J090033.49+421546.8,
J092914.49+282529.1 and J095937.11+131215.4).  The latter two have
very large relative velocities from the QSO (\dv\ $>$ 9500 \kms).  We
inspected the metal lines of J082638.59+515233.2, J090033.49+421546.8
(\dv\ = 820 and 3477 \kms\ respectively) in the SDSS spectra and did
not find them to be abnormally low.  High luminosities and moderate
relative velocities do therefore not necessarily lead to low
metallicities.

\section{Summary}

We have used a sample of 85 proximate absorbers in the SDSS with \dv\
$<$ 10,000 \kms\ and log \nhi\ $\ge$ 20 to investigate trends of metal
line strength with velocity separation.  Composite spectra are
constructed for 3 \dv\ ranges: \dv $<$ 3000 \kms, 3000 $<$ \dv\ $<$ 6000
\kms\ and \dv\ $>$ 6000 \kms.  The samples are further divided
according to \hi\ column density and QSO luminosity.  Our main findings are
as follows:

\begin{enumerate}

\item  Metal line EWs are largest in the \dv\ $<$ 3000 \kms\
composite and smallest at \dv\ $>$ 6000 \kms.  At intermediate
velocities (3000 $<$ \dv\ $<$ 6000 \kms) the EWs are between
the low and high \dv\ composites.  We interpret this result as
caused by higher metallicities at lower relative velocities.

\item Although the intermediate velocity composite indicates
that enhanced metallicities might exist out to several thousands
of \kms\ from the systemic QSO redshift, measurements of SiII $\lambda$
1526 in individual spectra indicate that the metallicity is most
enhanced at \dv\ $<$ 0 \kms.

\item The difference between the EWs in the lowest (\dv\ $<$ 3000
\kms) and highest (\dv\ $>$ 6000 \kms) composites is largest when
only the high \nhi\ (log \nhi\ $\ge$ 20.7) are considered.  Using
the SiII $\lambda$ 1526 EW-metallicity relation of Prochaska et al.
(2008a), we determine metallicities of 1/12 and 1/45 for the low
and high \dv\ composites respectively, when \nhi\ $\ge\ 20.7$.

\item  The absorbers within 3000 \kms\ of high luminosity QSOs
have lower metal line EWs than those within 3000 \kms\ of low
luminosity QSOs.  We speculate that this may be due to truncation
of star formation in the vicinity of a more luminous AGN.  However,
a low \dv\ and high QSO luminosity does not necessarily lead to
lower EWs.

\item Using the classifications of Herbert-Fort et al. (2006) we have
also shown that PDLAs are 5 times more likely to be `metal-strong'
than intervening DLAs.  We present 4 PDLAs whose SiII $\lambda$ 1808
is strong enough to be be detected even in the SDSS spectra which are
also classified as MSDLAs, making them excellent candidates for
follow-up of rare species.

\item No \lya\ emission is detected in the PDLA trough of the composite.
Although an accurate flux limit is difficult to model based on the
uncertainty of line emission properties, our simulations indicate
that typical \lya\ luminosities of a few $\times 10^{42}$ would have
been detected.

\item From an analysis of the power law slopes of QSOs with PDLAs
compared to a matched control sample, we place an upper limit of 
\ebmv\ $<$ 0.014 (for an SMC
extinction curve) on the reddening caused by PDLAs.  This is less than
inferred from a similar analysis of MS DLAs, and a complementary
investigation of reddening in proximate MgII absorbers at $z \sim
1.5$.

\end{enumerate}

Our findings are consistent with measurements in a smaller sample (16
PDLAs) of high resolution spectra which find that high \nhi\ PDLAs are
more metal-rich than the intervening population by a factor of $\sim$
3 (Ellison et al. 2010).  Taken together, these results are evidence
that galaxies in the vicinity of QSOs not only differ from the
intervening population but are directly affected by the proximity of
a QSO.

\section*{Acknowledgments} 

SLE acknowledges support from an NSERC Discovery Grant and Discovery
Accelerator Supplement.  Thank you to Andrew Fox for useful suggestions
on an earlier draft of this paper.

Funding for the SDSS and SDSS-II has been provided 
by the Alfred P. Sloan Foundation, the Participating Institutions, 
the National Science Foundation, the U.S. Department of Energy, the 
National Aeronautics and Space Administration, the Japanese 
Monbukagakusho, the Max Planck Society, and the Higher Education 
Funding Council for England. The SDSS Web Site is http://www.sdss.org/.

The SDSS is managed by the Astrophysical Research Consortium for the 
Participating Institutions. The Participating Institutions are the 
American Museum of Natural History, Astrophysical Institute Potsdam, 
University of Basel, University of Cambridge, Case Western Reserve 
University, University of Chicago, Drexel University, Fermilab, the 
Institute for Advanced Study, the Japan Participation Group, 
Johns Hopkins University, the Joint Institute for Nuclear Astrophysics, 
the Kavli Institute for Particle Astrophysics and Cosmology, the 
Korean Scientist Group, the Chinese Academy of Sciences (LAMOST), 
Los Alamos National Laboratory, the Max-Planck-Institute for Astronomy 
(MPIA), the Max-Planck-Institute for Astrophysics (MPA), New Mexico 
State University, Ohio State University, University of Pittsburgh, 
University of Portsmouth, Princeton University, the United States 
Naval Observatory, and the University of Washington.

\begin{appendix}
  
\section{PDLAs in the SDSS sample}

In Figure \ref{dla_fig} we present the fits to the \lya\ absorption
of the 85 PDLAs in our sample.

\begin{figure*}
\centerline{\rotatebox{0}{\resizebox{16cm}{!}{\includegraphics{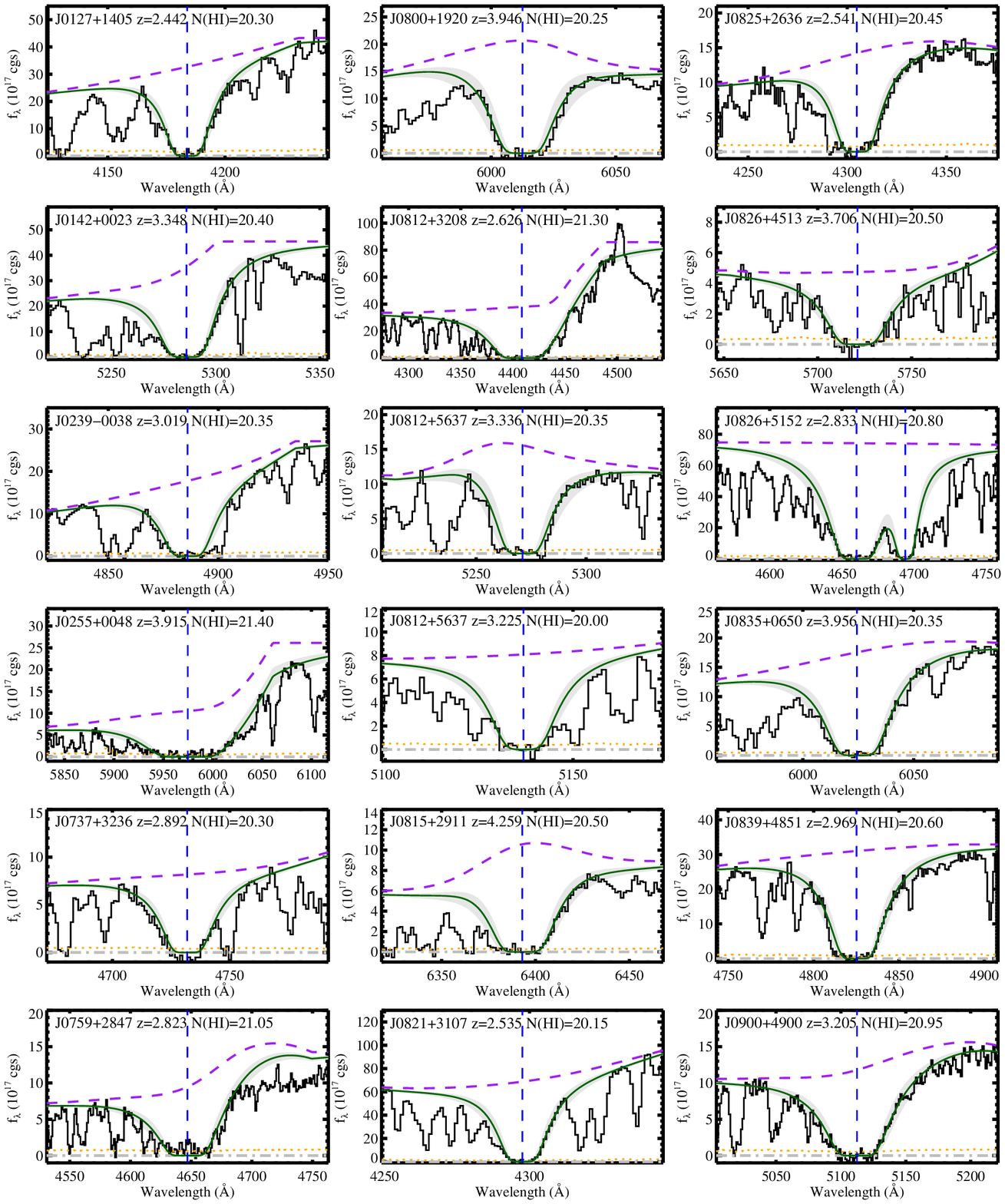}}}}
\caption{\label{dla_fig} Fits to the \lya\ absorption in our PDLA
sample, the blue vertical dashed line indicating the centre of the
\lya\ trough.  QSO names are abbreviated to Jhhmm+ddmm in order to fit
the panels.  Full SDSS QSO names are given in Table \ref{PDLA_tab}.
In cases where 2 PDLAs are present, they are shown in separate panels
but have been fitted simultaneously due to the significant overlap in
velocity space.  The grey shaded region shows the uncertainty in the
fit.  The grey dot-dashed line shows the zero level and orange dotted
line line the 1$\sigma$ error array.  The purple dashed line shows the
fit to the continuum.}
 \end{figure*}

\addtocounter{figure}{-1}
\begin{figure*}
\centerline{\rotatebox{0}{\resizebox{16cm}{!}{\includegraphics{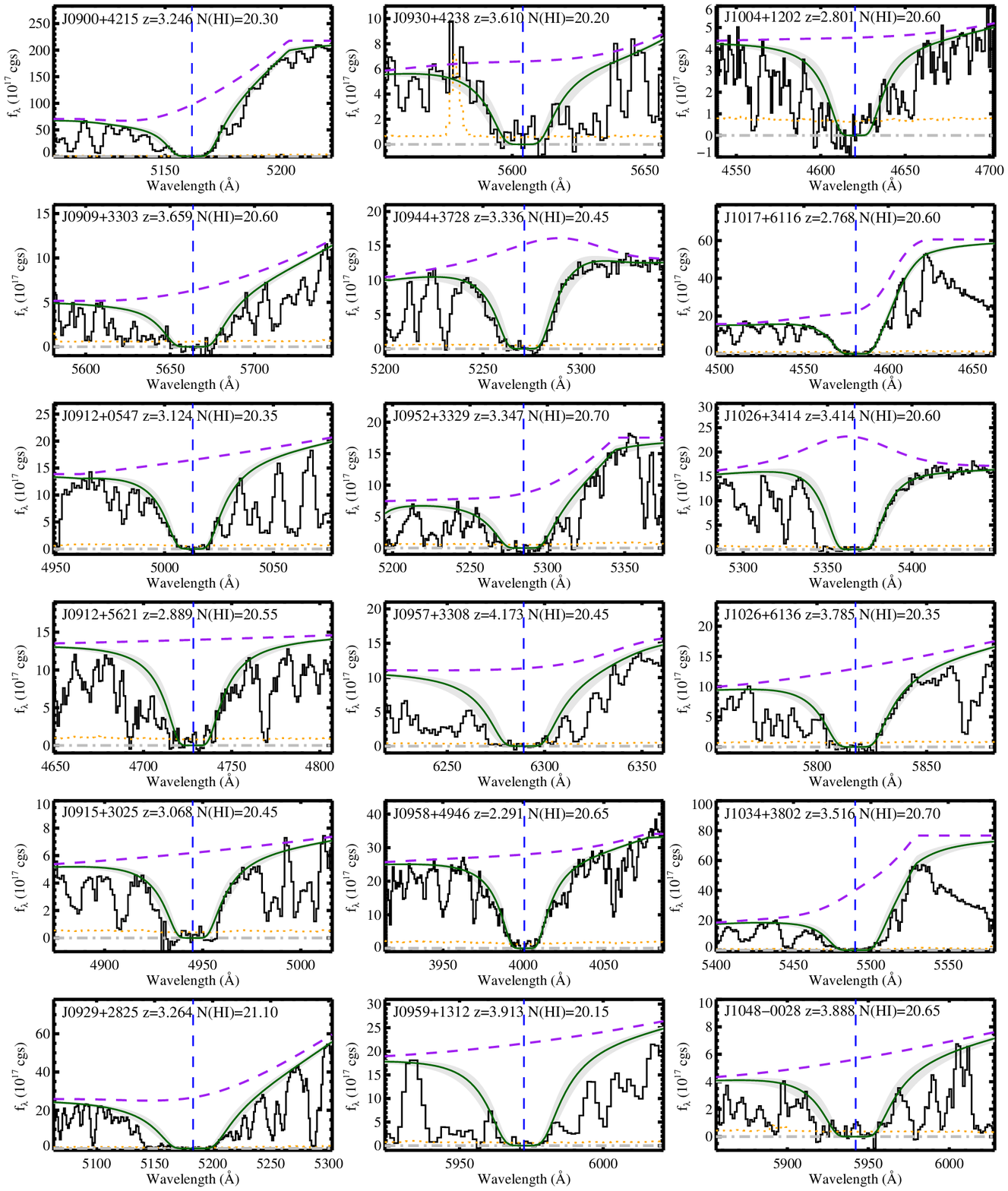}}}}
\caption{Continued.}
 \end{figure*}

\addtocounter{figure}{-1}
\begin{figure*}
\centerline{\rotatebox{0}{\resizebox{16cm}{!}{\includegraphics{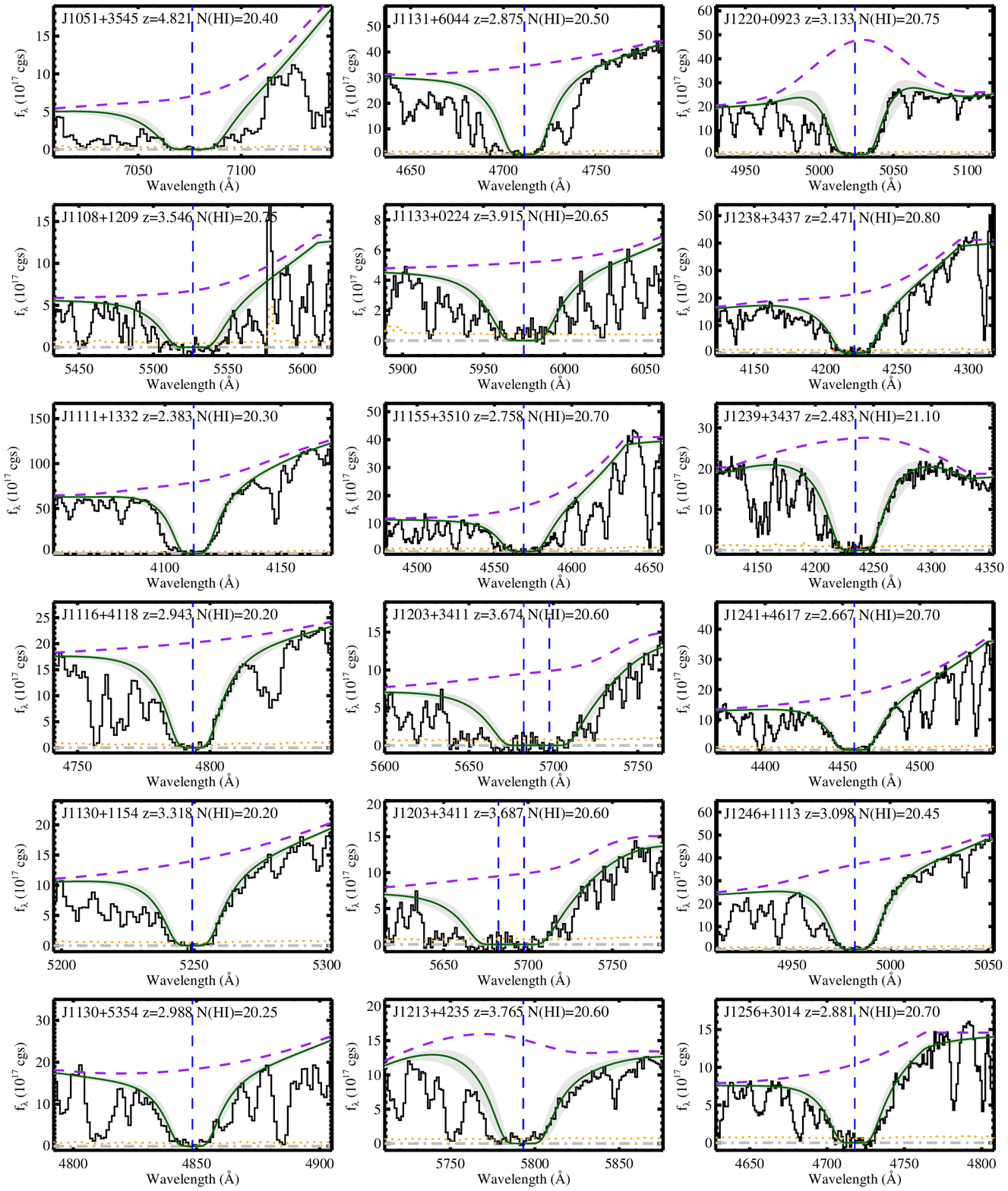}}}}
\caption{Continued.}
 \end{figure*}

\addtocounter{figure}{-1}
\begin{figure*}
\centerline{\rotatebox{0}{\resizebox{16cm}{!}{\includegraphics{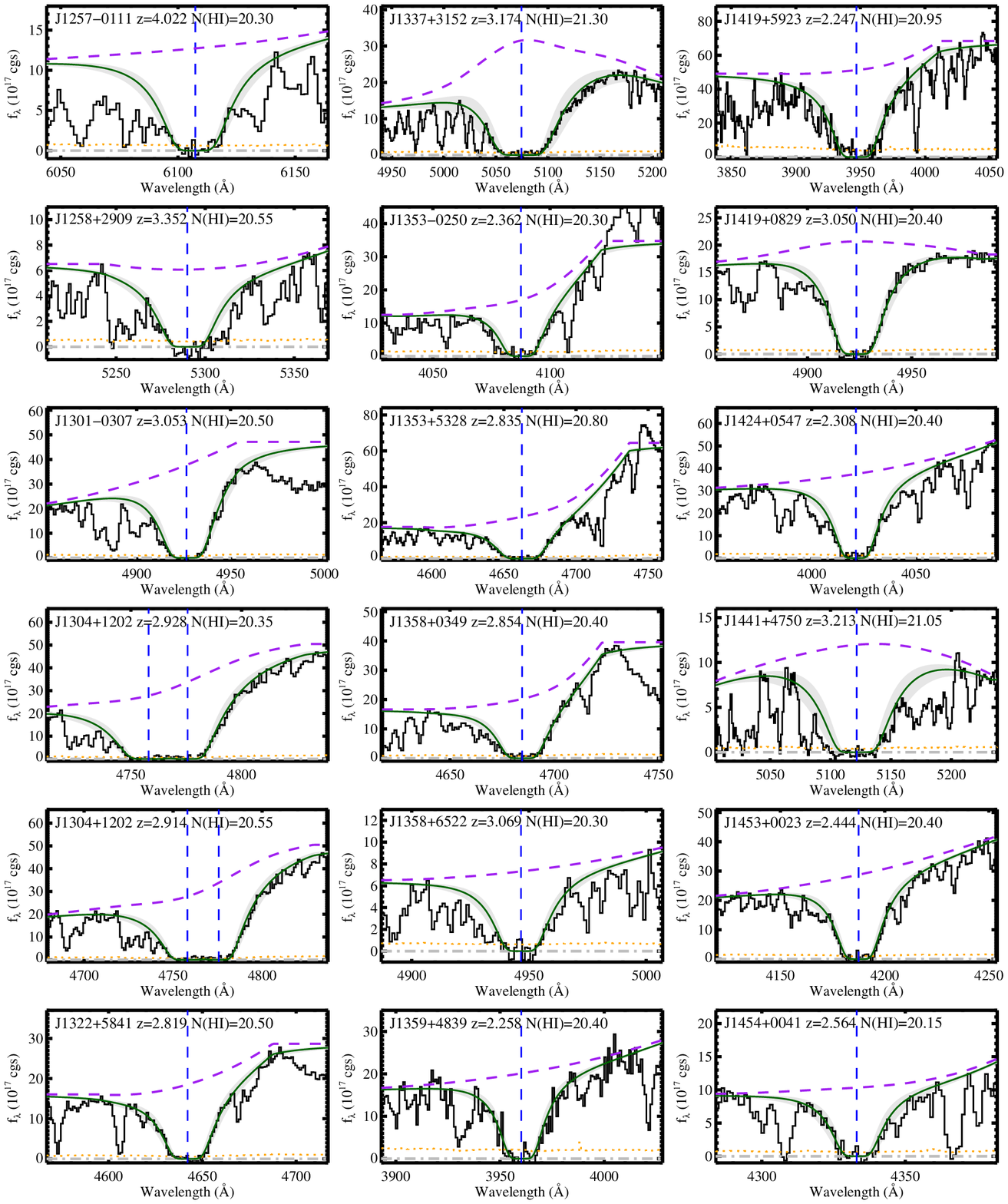}}}}
\caption{Continued.}
 \end{figure*}

\addtocounter{figure}{-1}
\begin{figure*}
\centerline{\rotatebox{0}{\resizebox{16cm}{!}{\includegraphics{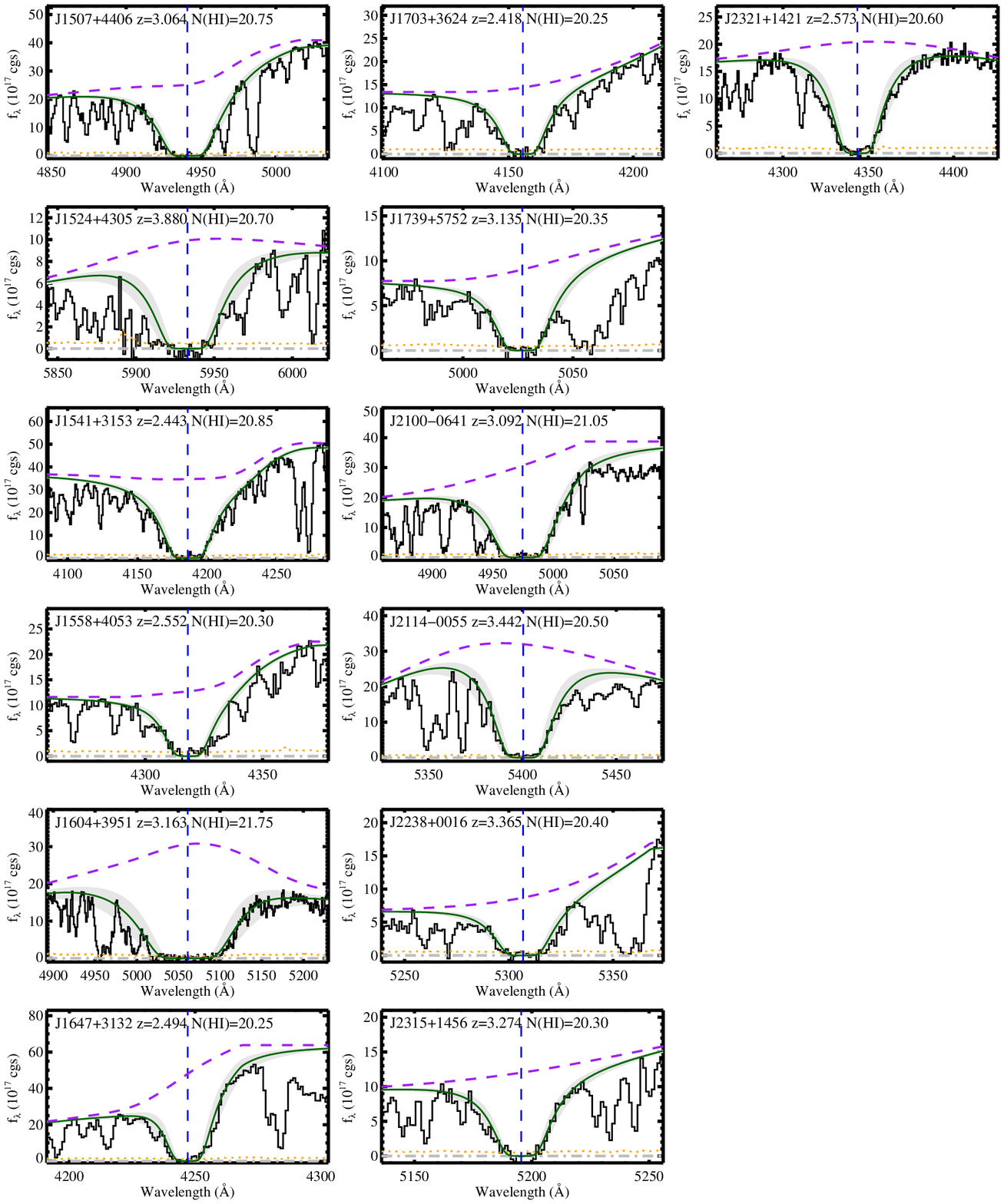}}}}
\caption{Continued.}
 \end{figure*}

\end{appendix}

\end{document}